\documentclass[iop]{emulateapj}

\usepackage{color}

\slugcomment {}

\shorttitle{HST Spectroscopy of WISE Brown Dwarfs}
\shortauthors{Schneider et al.}

\begin{document}

\title{Hubble Space Telescope Spectroscopy of Brown Dwarfs Discovered with the Wide-field Infrared Survey Explorer}

\author{Adam C. Schneider\altaffilmark{a}, Michael C. Cushing\altaffilmark{a}, J. Davy Kirkpatrick\altaffilmark{b}, Christopher R. Gelino\altaffilmark{b,c},  Gregory N. Mace\altaffilmark{d,e}, Edward L. Wright\altaffilmark{d}, Peter R. Eisenhardt\altaffilmark{f}, M. F. Skrutskie\altaffilmark{g}, Roger L. Griffith\altaffilmark{h}, \& Kenneth A. Marsh\altaffilmark{i}}   

\altaffiltext{a}{Department of Physics and Astronomy, University of Toledo, 2801 W. Bancroft St., Toledo, OH 43606, USA; Adam.Schneider@Utoledo.edu}
\altaffiltext{b}{Infrared Processing and Analysis Center, MS 100-22, California Institute of Technology, Pasadena, CA 91125, USA}
\altaffiltext{c}{NASA Exoplanet Science Institute, Mail Code 100-22, California Institute of Technology, 770 South Wilson Ave, Pasadena, CA 91125, USA}
\altaffiltext{d}{Department of Physics and Astronomy, UCLA, 430 Portola Plaza, Box 951547, Los Angeles, CA 90095-1547, USA}
\altaffiltext{e}{Department of Astronomy, The University of Texas at Austin, 2515 Speedway, Stop C1400, Austin, TX 78712, USA}
\altaffiltext{f}{Jet Propulsion Laboratory, California Institute of Technology, 4800 Oak Grove Dr., Pasadena, CA 91109, USA}
\altaffiltext{g}{Department of Astronomy, University of Virginia, 530 McCormick Road, Charlottesville, VA 22904, USA}
\altaffiltext{h}{Department of Astronomy \& Astrophysics, 525 Davey Lab, The Pennsylvania State University, University Park, PA 16802, USA}
\altaffiltext{i}{School of Physics and Astronomy, Cardiff University, Cardiff CF24 3AA, UK}

\begin{abstract}
We present a sample of brown dwarfs identified with the {\it Wide-field Infrared Survey Explorer} (WISE) for which we have obtained {\it Hubble Space Telescope} ({\it HST}) Wide Field Camera 3 (WFC3) near-infrared grism spectroscopy.  The sample (twenty-two in total) was observed with the G141 grism covering 1.10$-$1.70 $\mu$m, while fifteen were also observed with the G102 grism, which covers 0.90$-$1.10 $\mu$m.  The additional wavelength coverage provided by the G102 grism allows us to 1) search for spectroscopic features predicted to emerge at low effective temperatures (e.g.\ ammonia bands) and 2) construct a smooth spectral sequence across the T/Y boundary.  We find no evidence of absorption due to ammonia in the G102 spectra.  Six of these brown dwarfs are new discoveries, three of which are found to have spectral types of T8 or T9.  The remaining three, WISE J082507.35$+$280548.5 (Y0.5), WISE J120604.38$+$840110.6 (Y0), and WISE J235402.77$+$024015.0 (Y1) are the nineteenth, twentieth, and twenty-first spectroscopically confirmed Y dwarfs to date. We also present {\it HST} grism spectroscopy and reevaluate the spectral types of five brown dwarfs for which spectral types have been determined previously using other instruments.                   

\end{abstract}


\section{Introduction}
The {\it Wide-field Infrared Survey Explorer} (WISE) has been very successful at identifying the coolest brown dwarfs in the Solar neighborhood.  WISE provided all sky coverage at four mid-infrared wavelengths centered at 3.4, 4.6, 12 and 22 $\mu$m ($W1$, $W2$, $W3$, and $W4$).  This coverage has offered the ideal dataset with which to identify cool brown dwarfs whose spectral energy distributions peak at mid-infrared wavelengths.  Specifically, WISE was designed so that the $W1$ band coincides with a deep water$+$methane absorption feature and the $W2$ band coincides with a region largely free of opacity.  Thus, the WISE $W1-W2$ color has been especially useful for identifying late type dwarfs (\citealt{kirk11}, \citealt{cush11}, \citealt{kirk12}, \citealt{mace13}, and \citealt{thom13}).  Seventeen of the eighteen spectroscopically confirmed brown dwarfs with spectral types of Y0 or later were first identified by WISE (\citealt{cush11}, \citealt{kirk12}, \citealt{tin12}, \citealt{kirk13}, \citealt{cush14a}, and \citealt{pin14}).   The Y0 dwarf WISE J1217$+$16B was identified as a companion to a T8.5 dwarf, itself identified by WISE (\citealt{liu12}, \citealt{leg14}).  Three objects, WD 0806$-$661\citep{luh11}, CFBDSIR J1458$+$1013B \citep{liu11}, and WISE 0855$-$0714 \citep{luh14}, likely have effective temperatures similar to (or less than) the above Y dwarfs, but have yet to be spectroscopically confirmed.    

Ground-based follow-up observations at the mid-infrared wavelengths where the spectral energy distributions of cold brown dwarfs peak are nearly impossible due to the high thermal background, forcing follow-up observations to shorter wavelengths where these dwarfs are extremely faint.  As a result, obtaining moderate signal-to-noise (S/N) near-infrared spectra for many of the coldest WISE candidates has only been capable with the Wide Field Camera 3 (WFC3) aboard the {\it Hubble Space Telescope} ({\it HST}).  {\it HST} grism spectroscopy of the latest type brown dwarfs has been invaluable in the study and classification of these objects (\citealt{cush11}, \citealt{kirk12}, \citealt{kirk13}, and \citealt{cush14a}).  

As effective temperatures cool below 600 K, several spectroscopic features are predicted by model atmospheres to arise in the Y-band spectral region around 1.07 $\mu$m.  These include the emergence of ammonia absorption components and the disappearance of optical alkali resonance lines which have broad wings that are predicted to extend into the near-infrared (\citealt{bur00}, \citealt{kirk12}).  ({\it HST}) WFC3 G102 spectroscopy (0.90$-$1.10 $\mu$m)  allows us to investigate this additional wavelength range for such features, as well as inspect for differences (and similarities) as a function of spectral type. 

In this paper, we present our {\it HST} brown dwarf spectroscopic sample, including three new Y-dwarfs: WISEA J082507.37$+$280548.2, WISEA J120604.25$+$840110.5, and WISEA J235402.79$+$024014.1.  We also present improved (higher S/N) {\it HST} spectroscopy of five additional brown dwarfs for which near-infrared spectroscopy has been published previously, and present {\it HST}  G102 grism spectroscopy for a sample of 15 late T and Y dwarfs.  We first present the six new WISE brown dwarf discoveries, then present the new {\it HST} WFC3 spectroscopy of previously identified brown dwarfs.  We then present and analyze the Y-band spectra for the 15 brown dwarfs observed with the G102 grism and construct a complete spectral sequence across the T/Y boundary.  Lastly, we estimate physical properties of our entire sample of brown dwarfs by atmospheric model fitting. 

\section{The Sample}
The selection criteria for brown dwarfs in this study are described in detail in \cite{kirk12}.  AllWISE source catalog positions and photometry for all 22 brown dwarfs in this study including the six new discoveries are given in Table 1.  (Note: A detailed analysis of the {\it HST} spectrum of WISEA J182831.08$+$265037.6 will be published separately in Cushing et al., in preparation).  We provide data from the AllWISE catalog  because of its improved astrometric accuracy and increased sensitivity in the W1 and W2 bands compared to the WISE All-Sky catalog\footnote{See http://wise2.ipac.caltech.edu/docs/release/allwise/expsup/}.  An in depth characterization of the AllWISE source catalog can be found in \cite{kirk14} and the AllWISE Explanatory Supplement$^{10}$.  Hereafter, source names are abbreviated using the first four digits of the right ascension and declination for each object (e.g.\  WISEA J120604.25$+$840110.5 is WISE 1206$+$8401).  Finder charts for each of the new discoveries are presented in Figures 1$-$6.       

\begin{figure}
\plotone{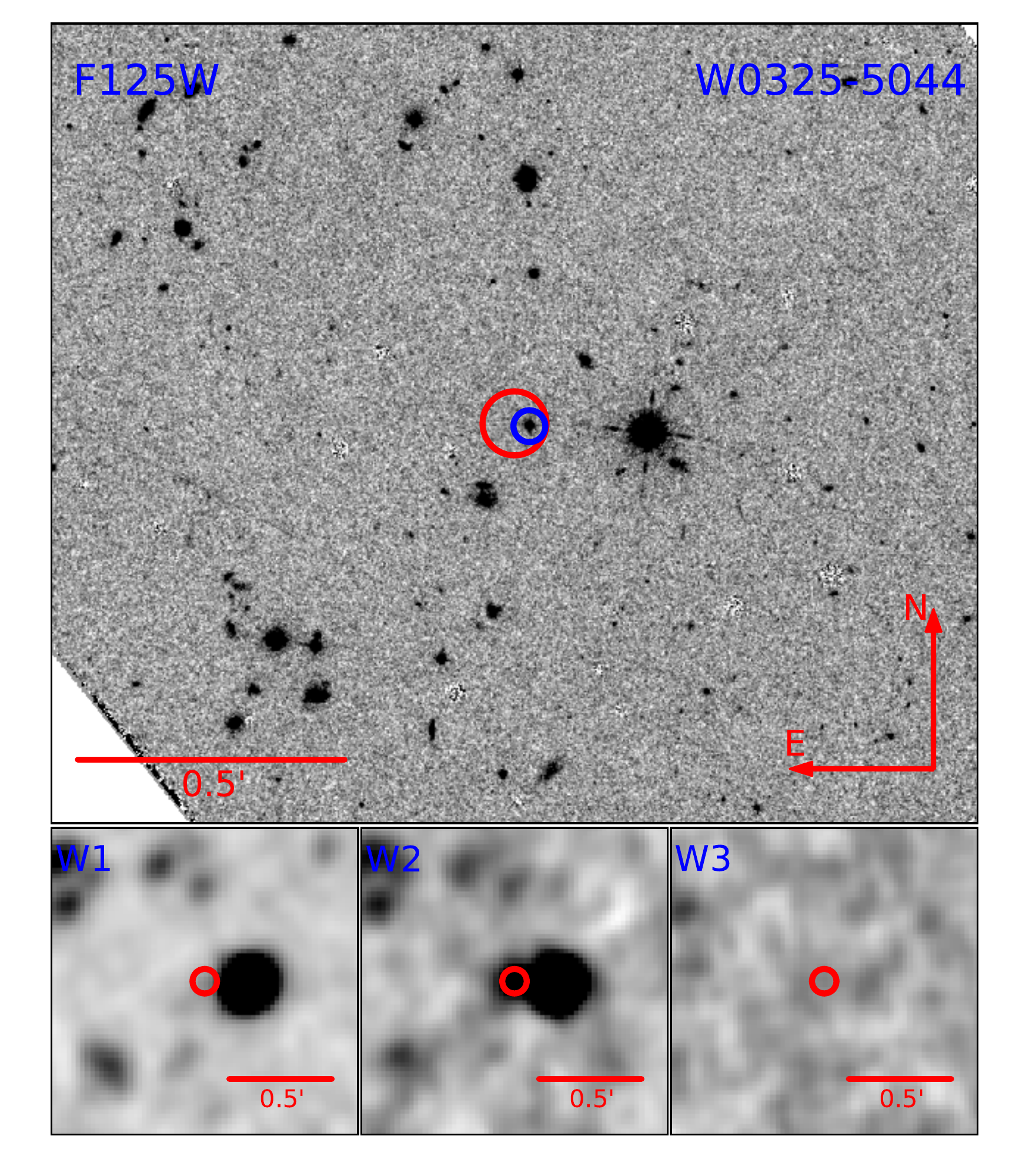}
\caption{Finder chart for new brown dwarf discovery WISE 0325$-$5044.  The top panel is the {\it HST} WFC3 F125W image, while the bottom three panels are WISE channels 1$-$3, from left to right.  All panels are centered on the WISE position of the brown dwarf, which is also indicated by a red circle.  In the top panel, the {\it HST} brown dwarf position is indicated by a blue circle.  North is up and East is left in each panel.}  
\end{figure}

\begin{figure}
\plotone{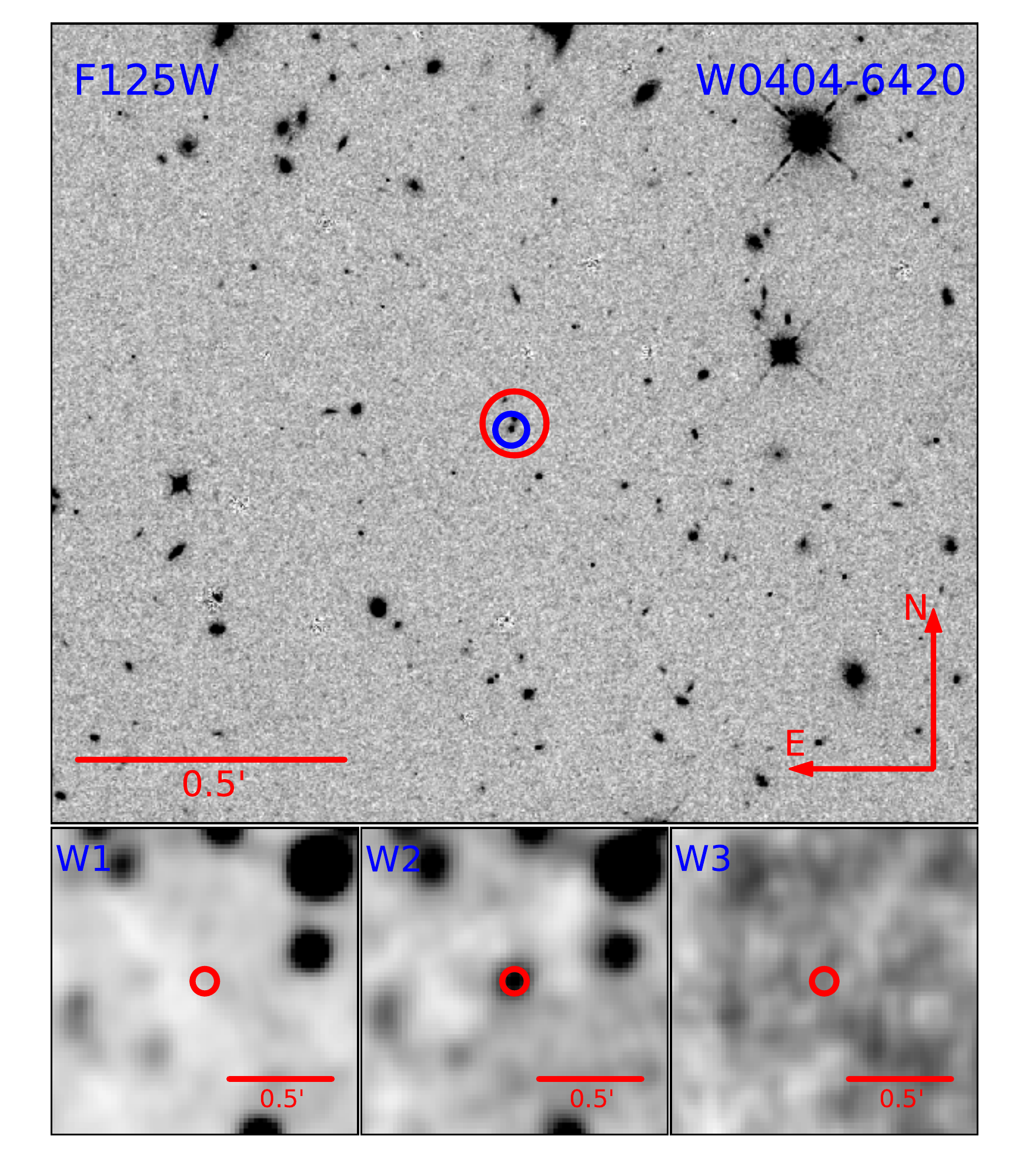}
\caption{Same as Fig.\ 1 for WISE 0404$-$6420.}
\end{figure}

\begin{figure}
\plotone{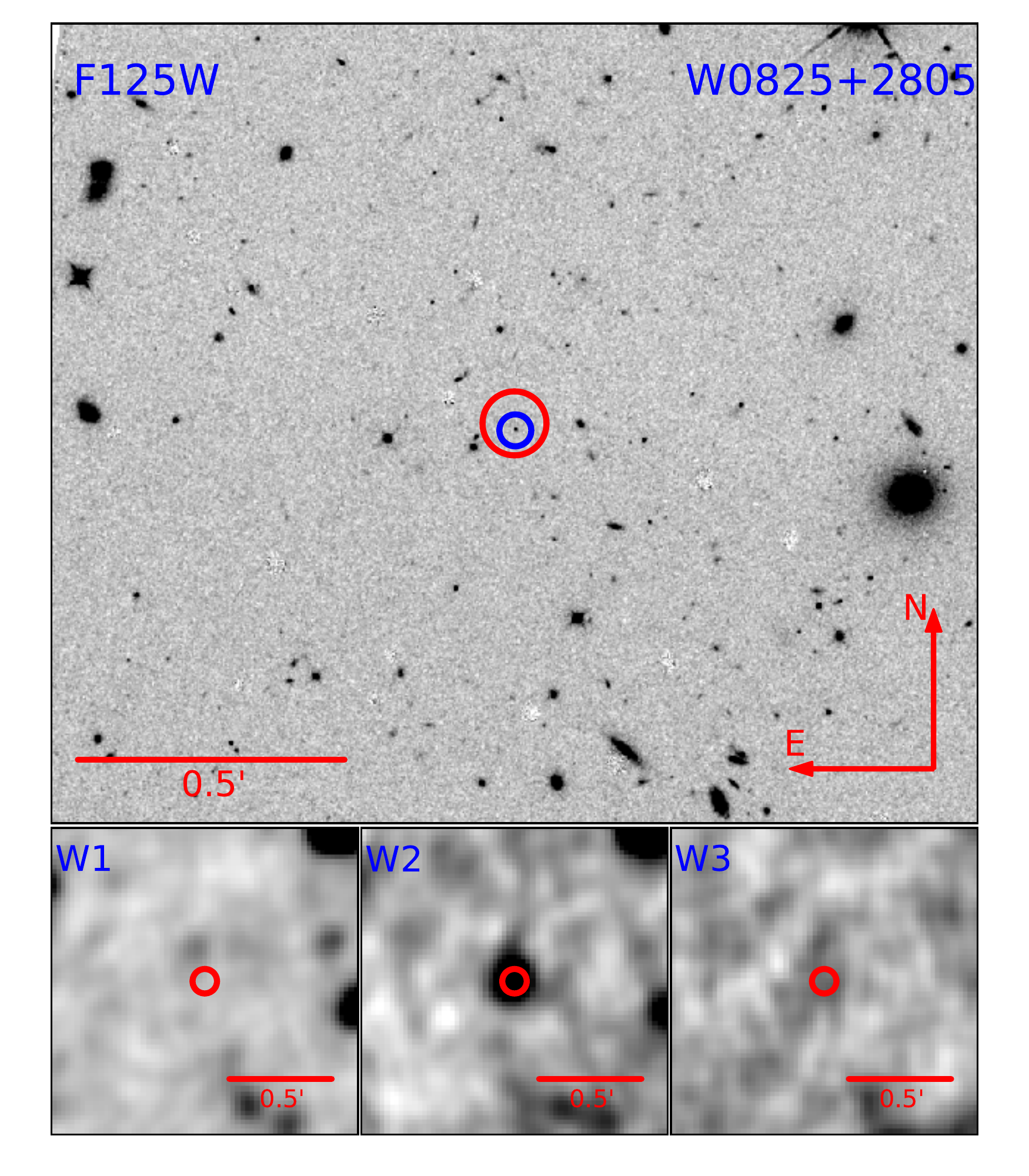}
\caption{Same as Fig.\ 1 for WISE 0825$+$2805.}
\end{figure}

\begin{figure}
\plotone{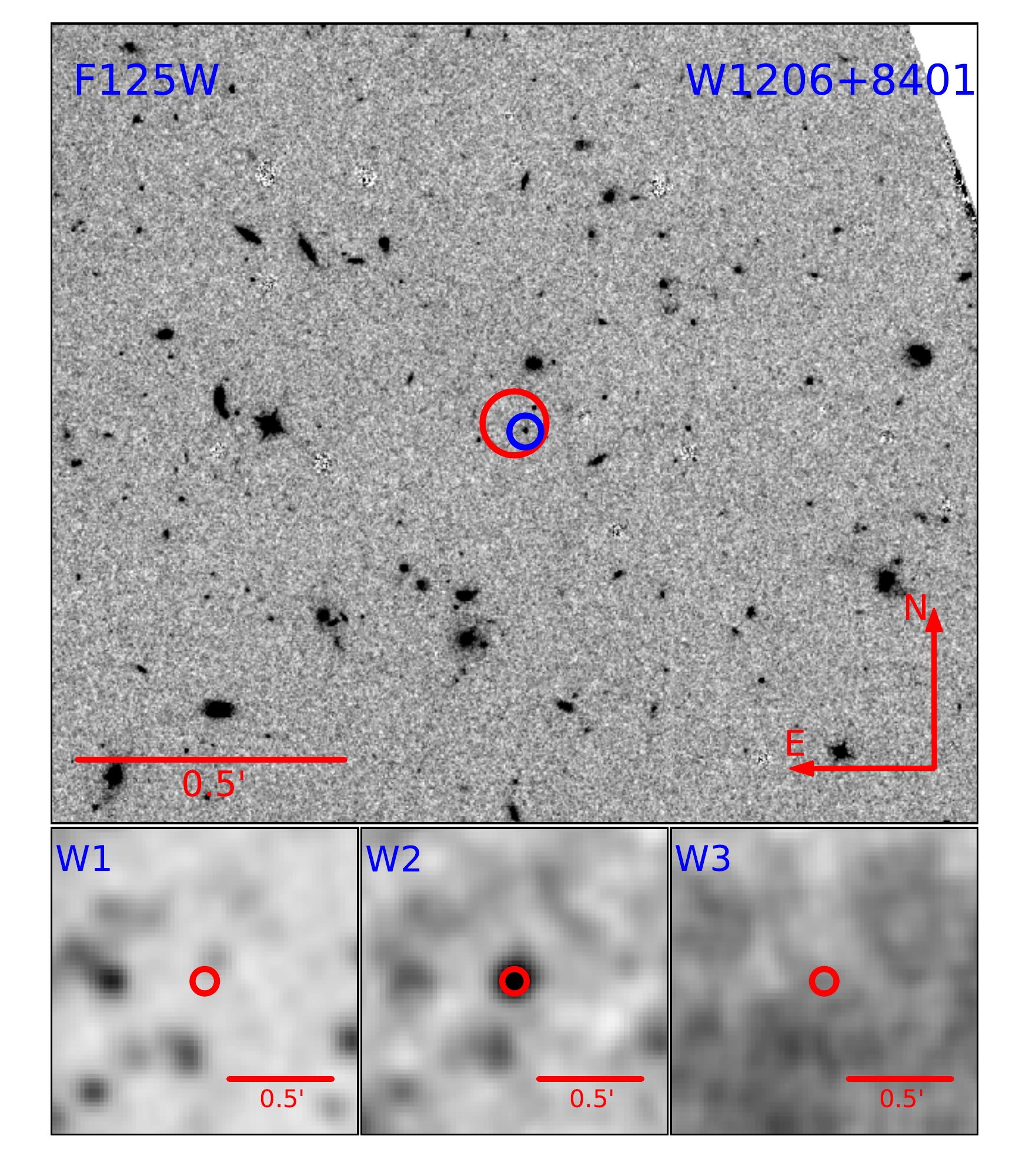}
\caption{Same as Fig.\ 1 for WISE 1206$+$8401.}
\end{figure}

\begin{figure}
\plotone{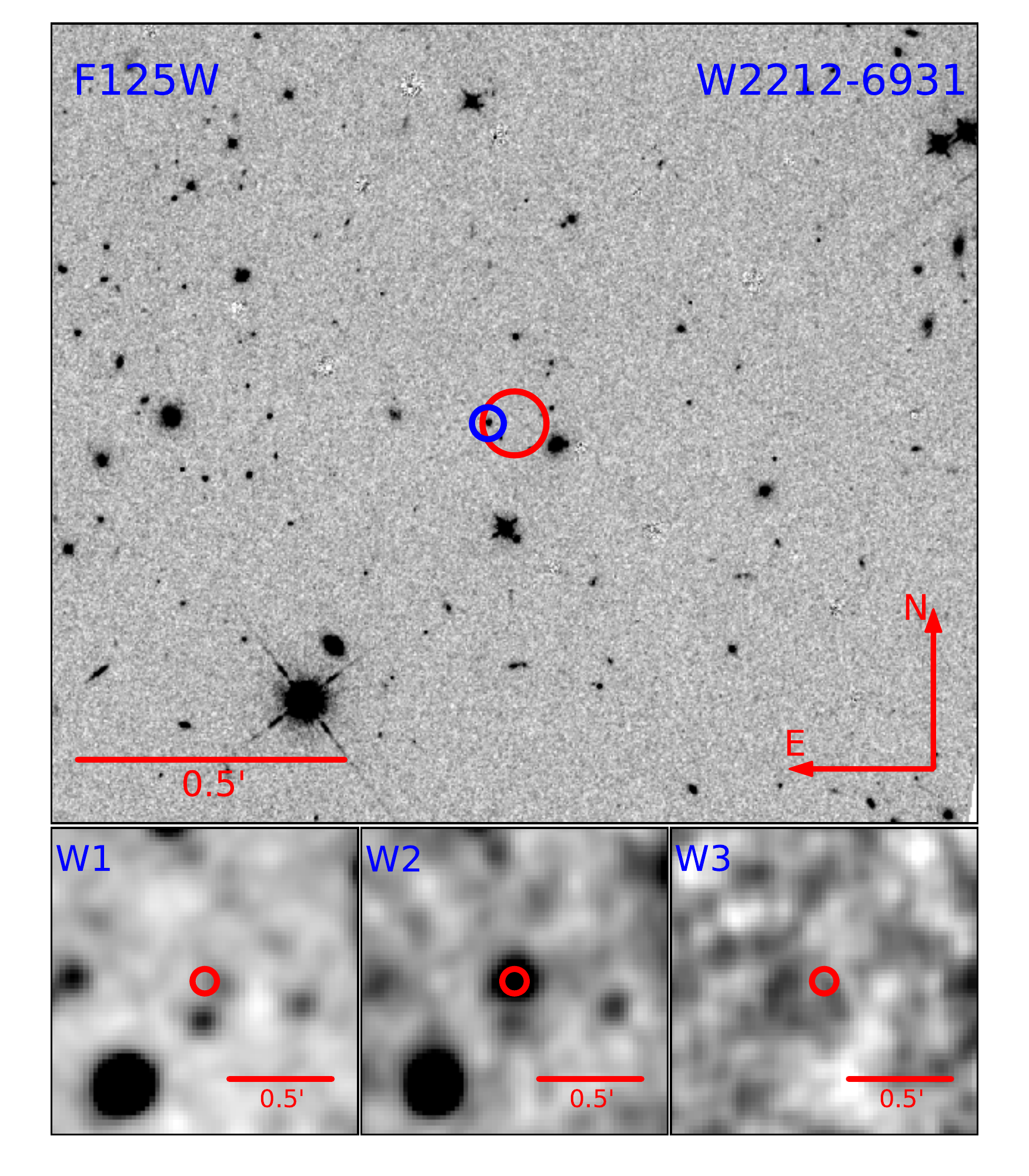}
\caption{Same as Fig.\ 1 for WISE 2212$-$6931.}
\end{figure}

\begin{figure}
\plotone{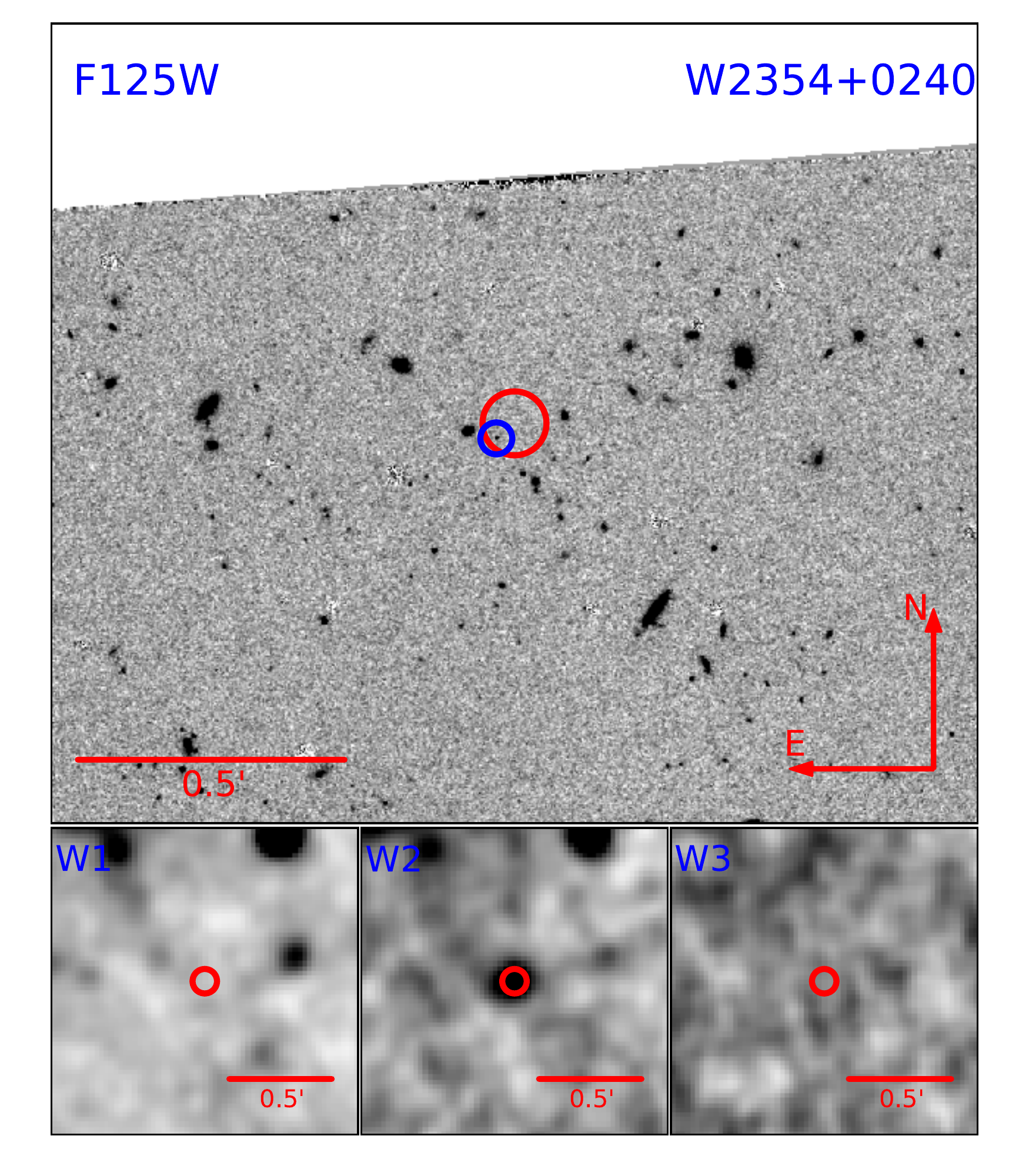}
\caption{Same as Fig.\ 1 for WISE 2354$+$0240.}
\end{figure}

\section{Observations}

\subsection{Photometry}
\subsubsection{WFC3/HST}
Direct images of each brown dwarf were obtained with the Wide Field Camera 3 (WFC3; \citealt{kim08}) aboard the {\it Hubble Space Telescope}.  Each grism observation (Section 3.2) requires an accompanying direct image, necessary for locating sources and determining source sizes.  Positions and source sizes are then used to determine the placement and size of the corresponding grism extraction apertures as well as the wavelength zero-point.  These images can also be used for photometric purposes.  Direct images for the G141 grism observations were obtained with either the F140W ($\lambda_p$ = 1392.3 nm) or F125W filters ($\lambda_p$ = 1248.6 nm), while images for the G102 grism observations were obtained with the F105W filter ($\lambda_p$ = 1055.2 nm), where $\lambda_p$ is the ``pivot wavelength'' (see \citealt{tok05}).  The 1024 $\times$ 1024 HgCdTe detector used by WFC3 has a plate scale of $\sim$0\farcs13 per pixel, resulting in a total field of view of 123$\arcsec$ $\times$ 126$\arcsec$.  Multiple images were obtained for each source and were combined using AstroDrizzle \citep{gonz12}.    

To test the consistency of our measured magnitudes, we performed aperture photometry on both the individual flat field frames ({\it flt} - multiplied by the WFC3 IR pixel area map to account for geometric distortion) and the final drizzled image created by AstroDrizzle ({\it drz}).  In several instances, we found differences between the magnitudes derived from the {\it flt} and {\it drz} images of up to $\sim$0.2 mag, similar to the results of \cite{kirk12}.  Further investigation revealed the cause of these discrepancies to be liberal default values for the cosmic ray rejection algorithm of AstroDrizzle.  On occasion, this algorithm will flag a central pixel of a source as a cosmic ray, thus reducing the total flux in the combined final {\it drz} image compared to the original {\it flt} images.  Because cosmic rays are rejected in the calwf3 calibration program used to create the {\it flt} images, we corrected this issue by first turning off the cosmic ray rejection algorithm when combining the {\it flt} frames with AstroDrizzle.  The {\it drz} images created by AstroDrizzle in this way were found to contain several artifacts (not cosmic rays) that are not present in the {\it drz} images created by AstroDrizzle when the cosmic ray rejection algorithm is turned on.  We remove these artifacts by employing a Laplacian edge detection algorithm \citep{vand01} to the {\it drz} images created by AstroDrizzle with the cosmic ray rejection algorithm turned off.  Aperture photometry is then performed on these final images.   

One result of the drizzling process is that the noise in adjacent pixels is correlated.  For this reason, determining the rms noise in the local background using common background subtraction methods, such as a sky annulus, are inapplicable.  Instead, we estimate the background and its uncertainty by applying the same aperture used for the source photometry (0\farcs4) to 1000 random star-free positions (determined via a 3$\sigma$ clip) on each image.  We take the mean and standard deviation of these measurements as the background level and its uncertainty.  This uncertainty is added in quadrature to the signal uncertainty, determined by dividing source counts by the gain and total exposure time, to determine the final source uncertainty.  Magnitudes were calculated using the Vega system with zeropoints of 25.1845, 25.1439, and 25.4523 for F140W, F125W, and F105W images, respectively\footnote{http://www.stsci.edu/hst/wfc3/phot\_zp\_lbn}. {\it HST} photometric magnitudes and uncertainties are given in Table 2.

\subsubsection{IRAC/Spitzer}
The entire sample was observed with the InfraRed Array Camera (IRAC; \citealt{faz04}) aboard the {\it Spitzer Space Telescope}.  Each dwarf was imaged with IRAC channels 1 and 2 centered at 3.6 and 4.5 $\mu$m (hereafter, {\it ch1} and {\it ch2}).  A detailed description of the IRAC data reduction procedure used to determine {\it ch1} and {\it ch2} magnitudes is given in \cite{kirk11}.  {\it ch1} and {\it ch2} magnitudes for each dwarf are given in Table 2.  For WISE 1639$-$6847, which was blended with another source in its {\it ch1} image, only a {\it ch2} magnitude is given.

\subsection{Spectroscopy}

{\it HST} spectroscopic observations were carried out using the Wide Field Camera 3 (WFC3) as part of Cycle 18 program 12330 (PI: Kirkpatrick), Cycle 19 program 12544 (PI: Cushing) and Cycle 20 programs 12970 (PI: Cushing) and 13178 (PI: Kirkpatrick) and are summarized in Table 3.  The G141 (1.1$-$1.7 $\mu$m, $R$ $\tbond$ $\lambda$/$\Delta$$\lambda$ $\approx$ 130) and G102 (0.9$-$1.1 $\mu$m, $R$ $\approx$ 210) spectroscopic reductions were performed using the methods described in \cite{cush11}, with the exception of one final step, described in the following paragraphs.      

Because the G141 and G102 grism modes are slitless, source spectra are occasionally influenced by photons from a nearby source or uneven background fluctuations.  For this reason, we developed a custom source extraction routine that allows us to define source apertures and background regions on the individual stamp images of a spectrum produced with the AXEDRIZZLE routine (e.g.\ Figure 7).  Once an aperture is defined, we use the aperture corrections of \cite{kun11} and the appropriate grism sensitivity curves\footnote{http://www.stsci.edu/hst/wfc3/analysis/grism\_obs/wfc3-grism-resources.html} to flux calibrate the final spectrum.   There is also a 2\% absolute flux calibration uncertainty \citep{kun11} which we account for in all synthetic photometry calculations (Section 5.3) and model fits (Section 5.4).  An example of the functionality of our extraction technique is shown in Figure 7, where the effects of uneven background fluctuations are mitigated.  

Several sources had multiple {\it HST} visits (i.e. spectroscopic observations occurring on different dates).  In principle, direct and spectroscopic images from multiple visits can be combined using standard AstroDrizzle routines to produce final images with improved S/N.  Yet, because our sources are nearby brown dwarfs, they tend to have significant proper motions, even on time scales as short as a month.  For {\it HST} grism observations, identical positions are critical for the placement of extraction apertures and wavelength zeropoints.  For sources with multiple visits, AXEDRIZZLE stamp images from each visit are median combined to produce a final spectroscopic image.  Spectra are then extracted using the method described above.  Objects with both G141 and G102 data are stitched together at 1.1 $\mu$m with no scaling done between the two spectra.  The complete spectroscopic sample is shown in Figures 8 and 9.  We note here that the depressed H-band peak of WISE 0647$-$6232 seen in \cite{kirk13} does not appear in our reduced spectrum.  This is likely due to either the added signal of our spectrum or our improved spectral extraction technique.  WISE 0535$-$7500 is located in an extremely crowded field, and thus  retrieving a clean G102 spectrum was particularly difficult, even with multiple observations at multiple roll angles.  While we did manage to extract a G102 spectrum for this object, we urge caution in its interpretation, as it is still likely contaminated.  We also note that a different roll angle allowed us to extract a cleaner J-band spectrum than that found in \cite{kirk12}, though the zeroth order light from a nearby object fell directly in the H-band portion of WISE 0535$-$7500's spectrum.  For this reason, we do not use the H-band portion of this object's spectrum.            

Upon inspection of their spectra, two objects that met the selection criteria were found to not be brown dwarfs.  The spectra for both of these objects were flat and devoid of spectroscopic features and are likely extragalactic in nature.  The AllWISE positions and photometry for these contaminants is provided in Table 1.

\begin{figure}
\plotone{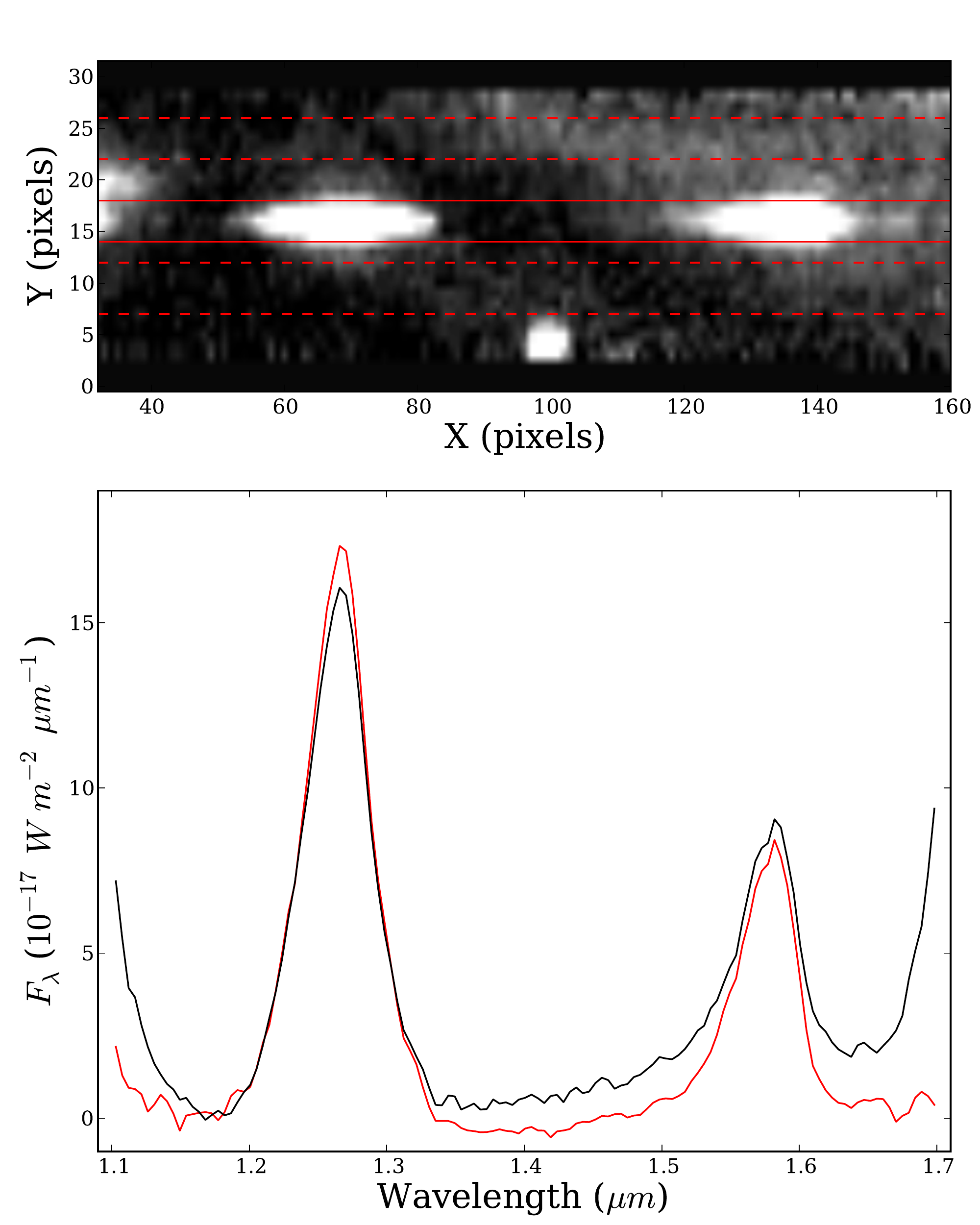}
\caption{{\it Top:} Stamp image of G141 the grism spectrum of WISE 2056$+$1459.  Solid red lines indicate the defined extraction aperture.  Dashed lines show the regions used for background fitting/subtraction.  {\it Bottom:} Comparison of the AXEDRIZZLE extracted spectrum of WISE 2056$+$1459 (black) and the spectrum extracted with our routine (red).}
\end{figure}

\begin{figure*}
\plotone{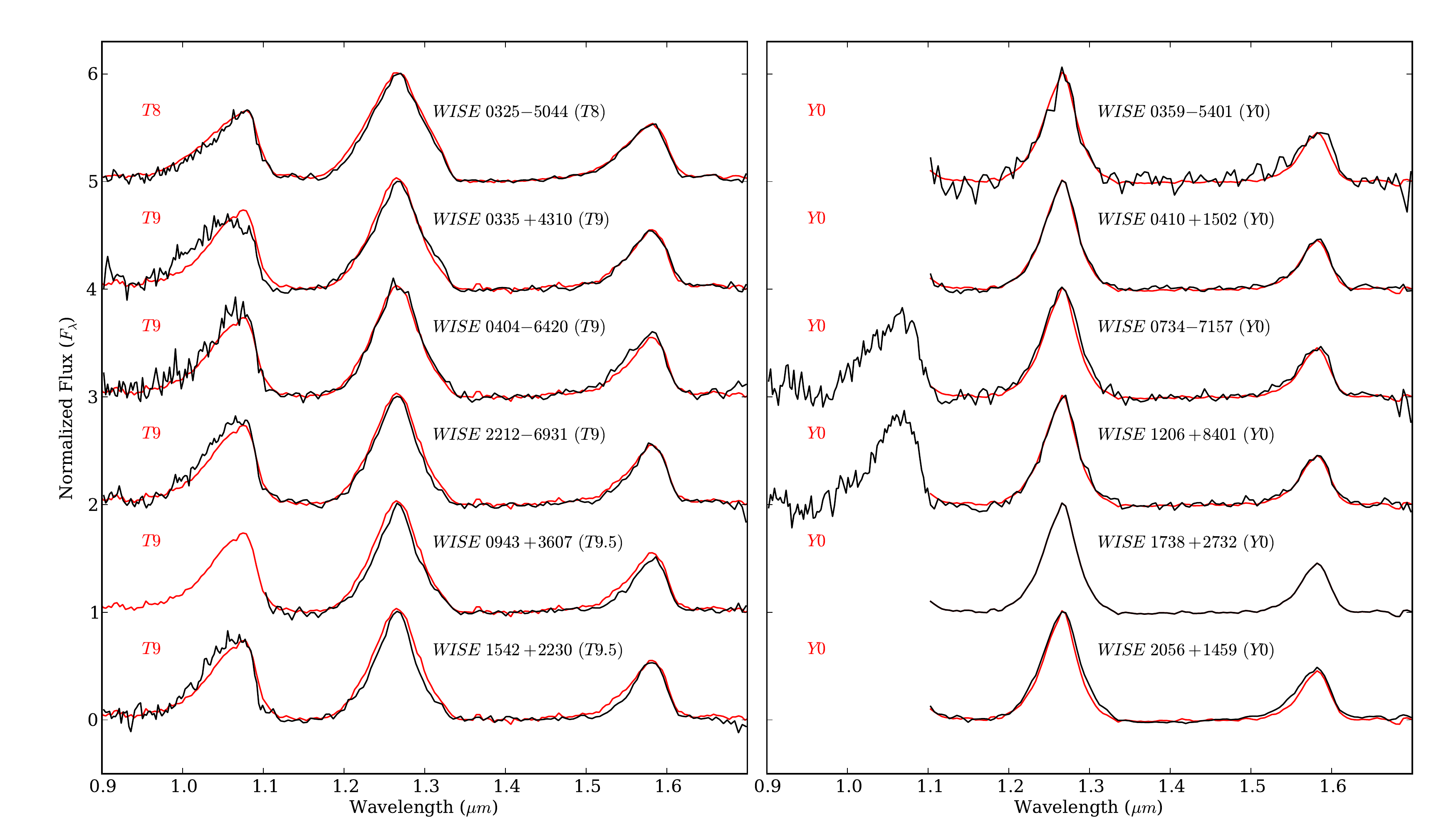}
\caption{All {\it HST} spectra (black) ordered by spectral type, normalized at 1.27 $\mu$m, offset by constants, and compared with corresponding spectral standards (red).  The spectral standards and the instruments from which the comparison spectra were obtained are as follows: T8$-$2MASSI J0415195$-$093506 (IRTF/SpeX; \citealt{burg06}), T9$-$UGPS J072227.51$-$054031.2 (IRTF/SpeX; \citealt{cush11}), Y0$-$WISE J173835.53$+$273259.0 (HST/WFC3; \citealt{cush11}), and Y1$-$WISE J035000.32$-$565830.2 (HST/WFC3; \citealt{kirk12}).}
\end{figure*}

\begin{figure*}
\plotone{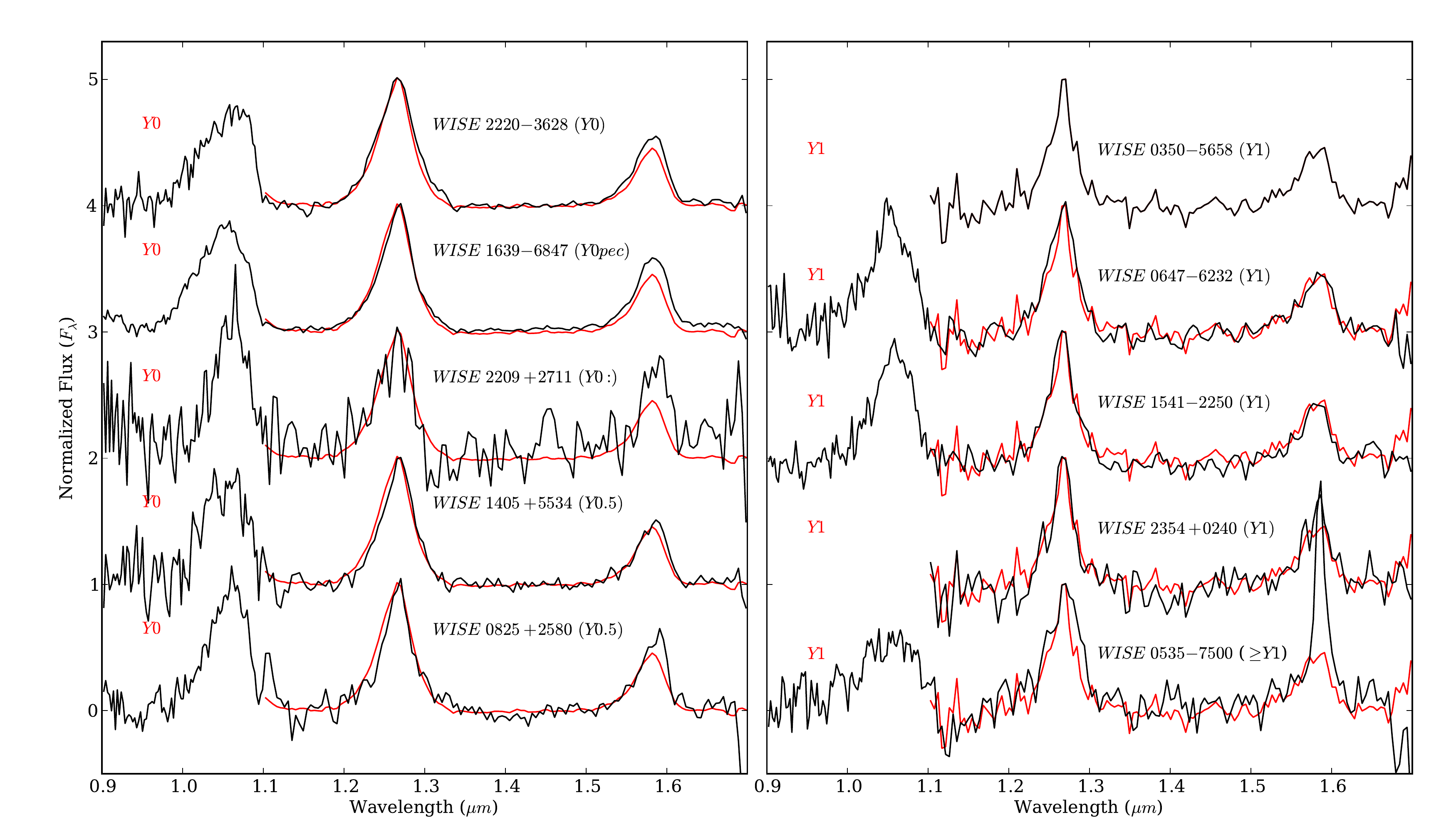}
\caption{All {\it HST} spectra (black) ordered by spectral type, normalized at 1.27 $\mu$m, offset by constants, and compared with corresponding spectral standards (red).  The spectral standards and the instruments from which the comparison spectra were obtained are as follows: T8$-$2MASSI J0415195$-$093506 (IRTF/SpeX; \citealt{burg06}), T9$-$UGPS J072227.51$-$054031.2 (IRTF/SpeX; \citealt{cush11}), Y0$-$WISE J173835.53$+$273259.0 (HST/WFC3; \citealt{cush11}), and Y1$-$WISE J035000.32$-$565830.2 (HST/WFC3; \citealt{kirk12}).}
\end{figure*}

\section{Analysis}

\subsection{New Discoveries}

Spectral types for the six new brown dwarf discoveries (WISE 0325$-$5044, WISE 0404$-$6420, WISE 0825$+$2805, WISE 1206$+$8401, WISE 2212$-$6931, and WISE 2354$+$0240) were determined following the methods of \cite{cush11} and \cite{kirk12}.  These discoveries are classified via by-eye comparisons of the width of the J-band peak to the spectral standards defined in \cite{burg06}, \cite{cush11}, and \cite{kirk13}.  Comparisons are shown in Figures 8 and 9, while the spectral types are provided in Table 1. 

The spectrum of WISE 0325$-$5044 is an excellent match to the T8 spectral standard 2MASSI J0415195$-$093506 \citep{burg06}, hence we classify it as a T8.  Similarly, WISE 0404$-$6420 and WISE 2212$-$6931 show excellent agreement with the T9 spectral standard UGPS J072227.51$-$054031.2 \citep{cush11} and are classified as such.  WISE 1206$+$8401 is an almost perfect match to the Y0 spectral standard WISE J173835.53$+$273259.0 \citep{cush11}, therefore we are confident in assigning it a Y0 spectral classification.  Even though the spectrum of WISE 2354$+$0240 is somewhat noisy, it shows a J-band peak significantly narrower than the Y0 spectral standard and matches well with the Y1 standard WISE J035000.32$-$565830.2 \citep{kirk12}.  Therefore, we assign a spectral type of Y1 for this object, making it the fourth such object to receive such a designation, along with WISE J035000.32$-$565830.2 \citep{kirk12}, WISE J064723.23$-$623235.5 \citep{kirk13}, and WISE J154151.65$-$225024.9 (see Section 4.2).  The most ambiguous of all classifications was that of WISE 0825$+$2805, which is shown separately in Figure 10.  A comparison of WISE 0825$+$2805 with the Y0 and Y1 spectral standards shows that the width of the J-band peak for this object is intermediate between the two.  For this reason we assign WISE 0825$+$2805 a spectral type of Y0.5.

While all of these discoveries are part of an extensive parallax program to measure their distances, we provide preliminary estimates here based solely on W2 photometry and the polynomial fits from \cite{dup12} and \cite{kirk12}.  Distances estimates are provide in Table 4.  Uncertainties are based solely on the photometric uncertainty of the W2 measurement.     

\begin{figure}
\plotone{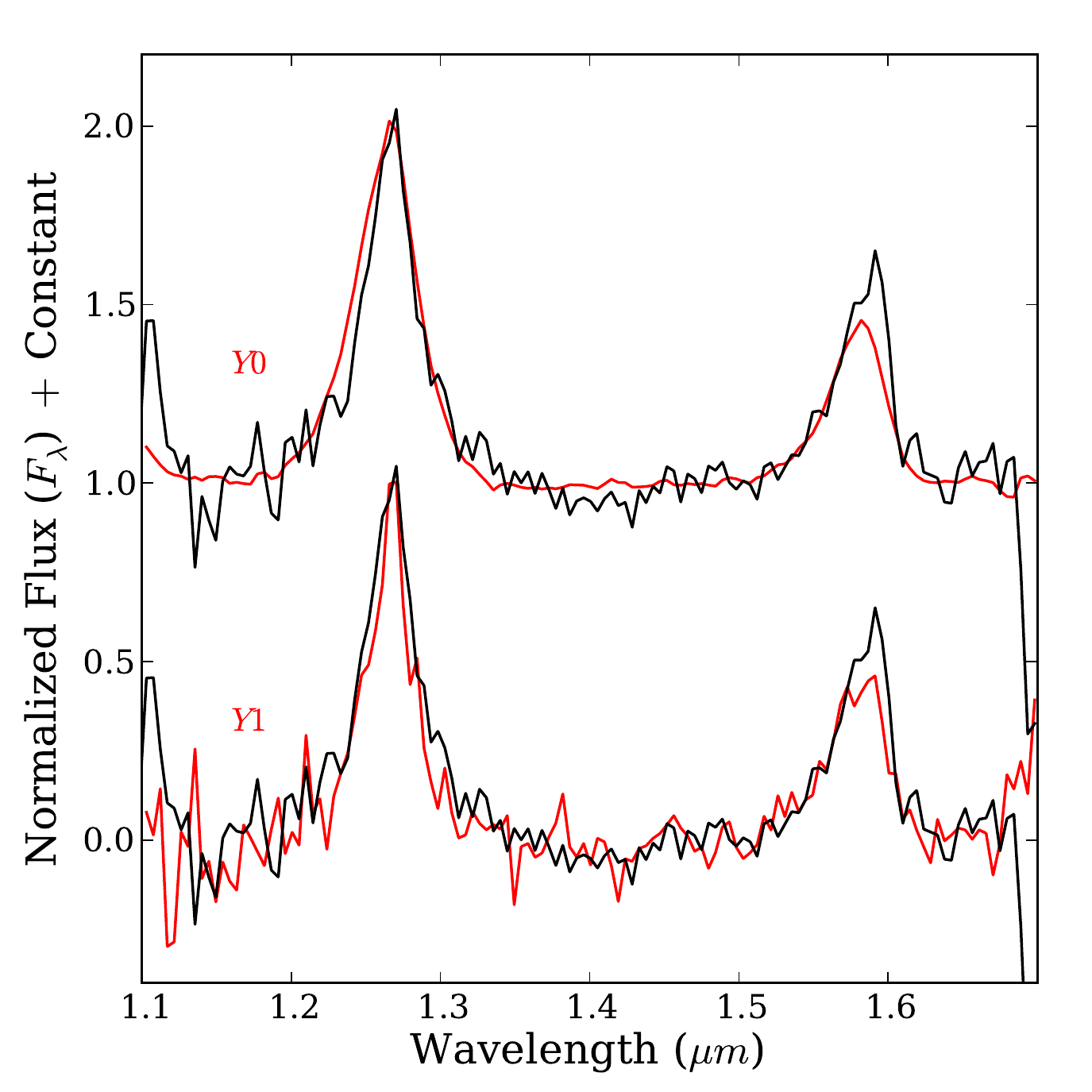}
\caption{Spectral classification of WISE 0825$+$2805.  WISE 0825$+$2805 is shown in black, while spectral standards are shown in red.}
\end{figure}

\subsection{Improved Spectroscopy of Previously Identified Brown Dwarfs}
Figures 8 and 9 also show {\it HST} spectroscopy of five previously known brown dwarfs, including WISE J0335$+$4310 (T9$-$\citealt{mace13}), WISE J07344$-$7157 (Y0$-$\citealt{kirk12}), WISE J1541$-$2250 (``Y0.5?''$-$\citealt{kirk12}), WISE J1639$-$6847 (Y0-Y0.5$-$\citealt{tin12}, and WISE J2220$-$3628 (Y0$-$\citealt{kirk12}). Spectral types were reexamined using the methods described in Section 4.1.  Spectral types determined in this way agree well with the spectral types determined in \cite{mace13}, \cite{kirk12}, and \cite{tin12}.  With the improved S/N from {\it HST}, we reclassify W1541$-$2250 as Y1 because the width of its J-band peak is a much better match to the Y1 standard than that of the Y0 standard.  This determination supersedes the ``Y0.5?''\ suggested in \cite{kirk12}.  The J-band peak of WISE 1639$-$6847 matches well with the Y0 standard, in agreement with the Y0-Y0.5 spectral type determined in \cite{tin12}.  However, the Y-band peak and $Y-J$ color of this object is unusual compared to other Y0 dwarfs (see Section 5.1 and Figure 13), therefore we classify WISE 1639$-$6847 as Y0pec.     

\section{Discussion}

\subsection{Spectral Sequence}

In Figure 11 we show the entire combined {\it HST} G102/G141 near infrared spectra for a complete sequence of late-type brown dwarfs from T8 to Y1.  There is a distinct trend for the Y, J, and H band peaks as temperatures decrease.  The Y band peak appears more symmetrical going from T8 to Y1.  This occurs because the blue wing of the Y-band peak becomes more suppressed for later spectral types.  The wavelength at which the peak reaches a maximum also occurs at a different location as temperature decreases, gradually becoming bluer with spectral type.  The J-band peak becomes narrower as temperatures decrease, while the H-band peak appears to become more symmetrical for later spectral types.  The peaks of both the J- and H-bands occur at the same position throughout the sequence.   

\begin{figure*}
\plotone{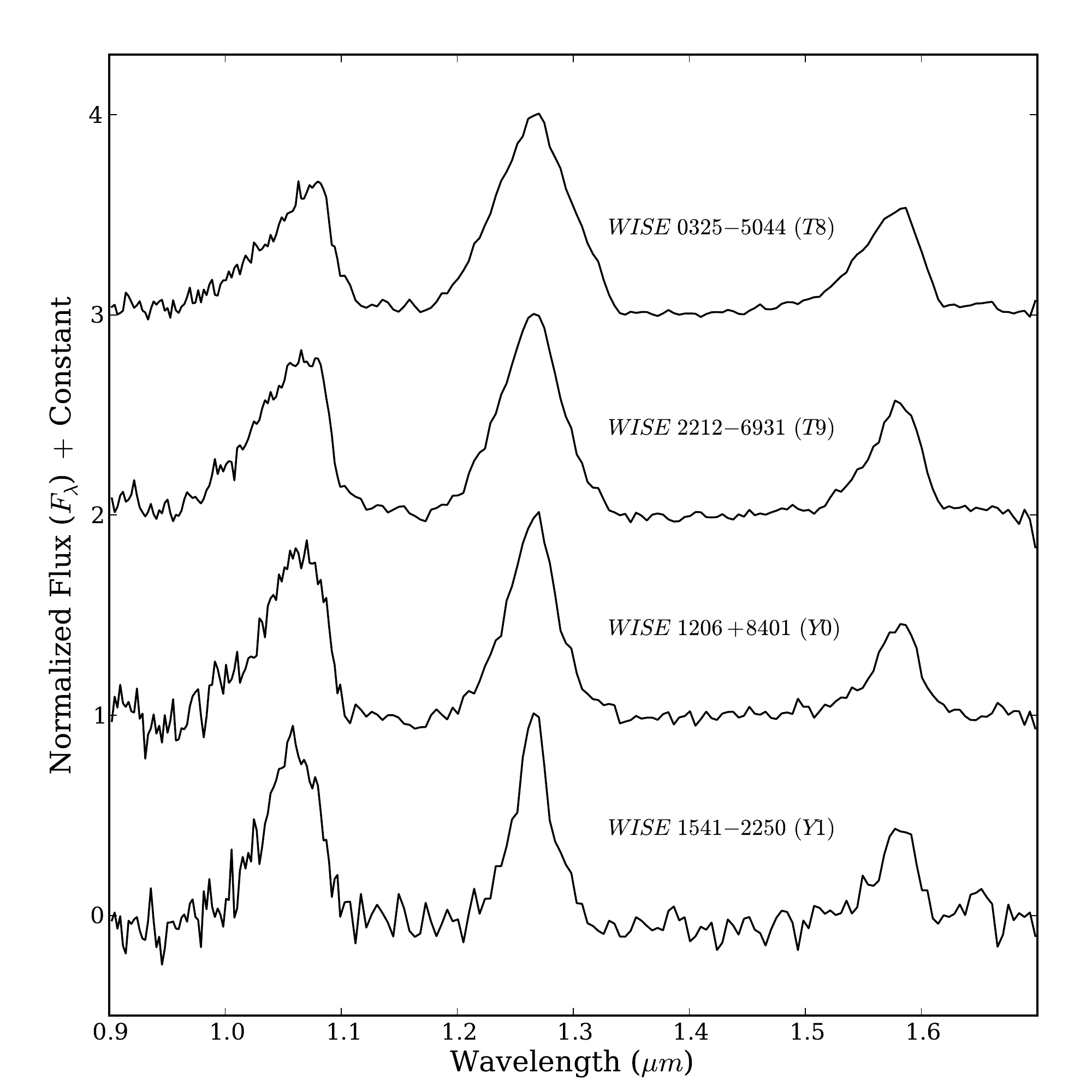}
\caption{The complete spectral sequence from T8 to Y1.}
\end{figure*}

G102 observations were made for five of the six new brown dwarfs presented in Section 4.1 (the exception being WISE 2354$+$0240), all five dwarfs presented in Section 4.2, and five additional brown dwarfs, including WISE 0535$-$7500 ($\geq$Y1$-$\citealt{kirk12}), WISE 0647$-$6232 (Y1$-$\citealt{kirk13}), WISE 1405$+$5534 (Y0pec$-$\citealt{cush11}), WISE 1542$+$2230 (T9.5$-$\citealt{mace13}), and WISE 2209$+$2711 (Y0:$-$\citealt{cush14a}).  The G102 spectra are presented in Figure 12. 

WISE 1639$-$6847 shows an increase in the relative height of the Y-band peak compared to other Y0 type dwarfs.  The peak is also shifted blueward from the peak of the T9 standard UGPS J072227.51$-$054031.2.  This is a surprising result considering that we find the J-band peak matches well with the Y0 standard (see Figure 9).  Because the Y-band shape of this object is so unusual, we classify it as a Y0pec.  We note that a similarly large Y-band peak is seen in the FIRE spectrum of WISE 1639$-$6847 presented in \cite{tin12} (see their Figure 7).  

The Y0pec dwarf WISE 1405$+$5534 (which is categorized as `pec' due to a wavelength shift of its H-band peak) shows a Y-band peak slightly different than that of the T9 and Y0 dwarfs.  Though noisy, the Y-band peak of WISE 1405$+$5534 is more symmetrical about its maximum than the normal T9-Y0 dwarfs.  We also note here that the J-band peak appears slightly narrower than that of the Y0 standard (see Figure 9).  For this reason, we reclassify WISE 1405$+$5534 as Y0.5.  The near-infrared colors of WISE 1405$+$5534 support this change (see Section 5.3).

WISE 2209$+$2711 (Y0:$-$\citealt{cush14a}), while noisy, shows a very peaked appearance, more similar to that of the Y1s than the Y0s.  This may be providing a hint that this object is possibly Y1 or intermediate between Y0 and Y1.

\begin{figure*}
\plotone{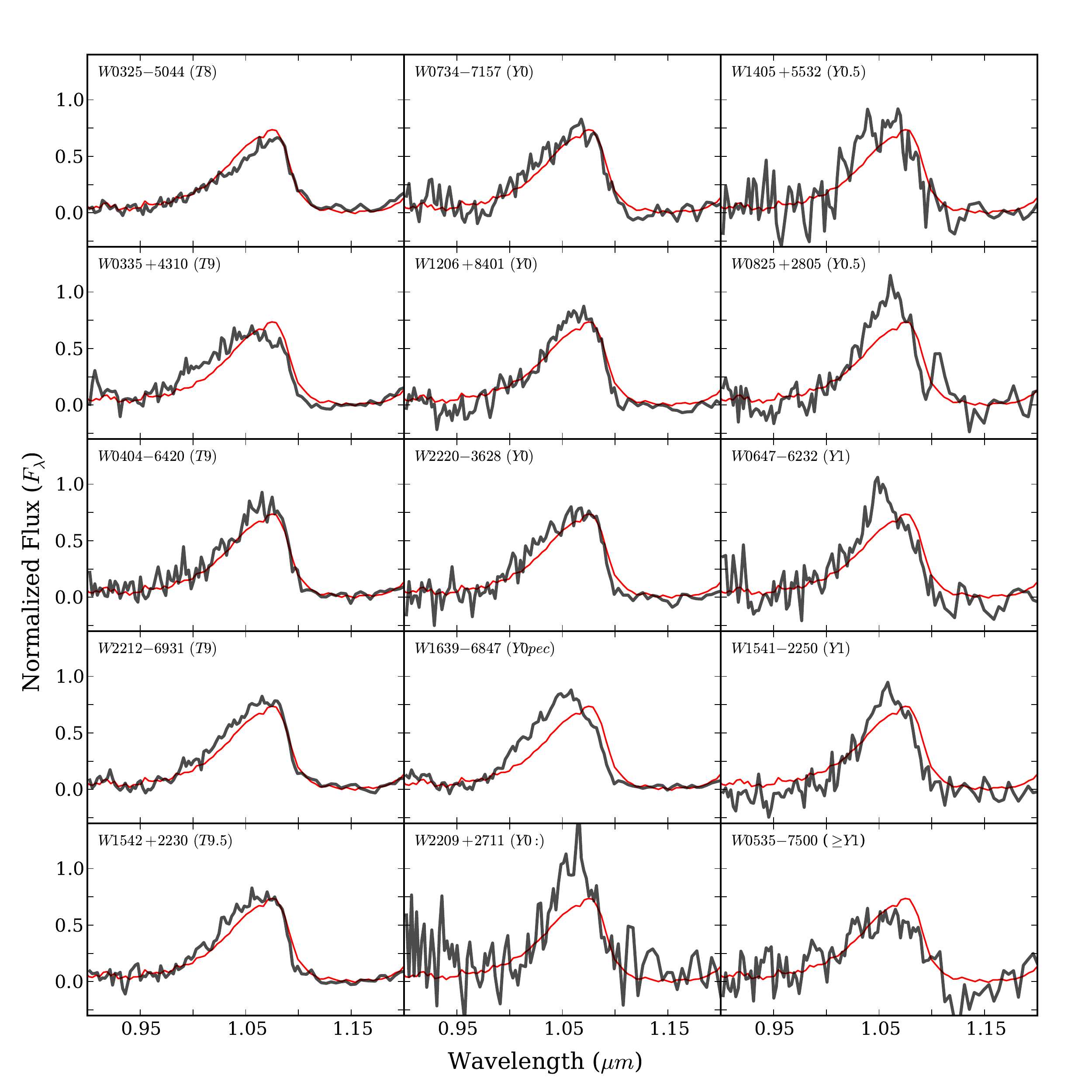}
\caption{{\it HST} G102 spectra for WISE brown dwarfs (black).  Each spectrum is normalized at 1.27 $\mu$m (the J-band peak) and ordered sequentially by spectral type from top to bottom, then left to right.  For comparison purposes, the spectrum of the T9 standard UGPS J072227.51$-$054031.2 is included in red on each plot. }
\end{figure*}

\subsection{$NH_3$}
The appearance of ammonia (NH$_3$) in the near infrared spectra of cool brown dwarfs has been suggested for some time as the harbinger of the Y spectral class (\citealt{bur03}, \citealt{leg07}, and \citealt{kirk08}).  Models predict that NH$_3$ absorption bands could become prominent in brown dwarfs with temperatures $\lesssim$450 K.  \cite{cush11} show a possible detection of NH$_3$ on the blue wing of the H-band peak of WISE 1738$+$2732, but this NH$_3$ feature is itself confused with a strong H$_2$O band.  An additional NH$_3$ absorption component is predicted to emerge that will affect the shape of the Y-band spectrum of extremely cool dwarfs, eventually causing a bifurcation of the peak around 300 K.  So far, a firm detection of NH$_3$ absorption at 1.03 $\mu$m has been elusive \citep{leg13}.  \cite{leg13} have suggested that vertical mixing can explain the lack of NH$_3$ in the spectrum of the Y0 dwarf WISE 2056$+$1459.  Vertical mixing keeps the atmospheric chemistry from reaching equilibrium, thus enhancing the N$_2$ abundance and decreasing the NH$_3$ abundance (\citealt{sau03}, \citealt{hub07}, \citealt{mor14}).  Now, with an increased sample size of late-type brown dwarfs for which Y-band spectra are available, we can investigate the effects of NH$_3$ in these atmospheres.  If the lack of an NH$_3$ detection in the spectrum of the Y0 dwarf WISE 2056$+$1459 is due to vertical mixing, as suggested by \cite{leg13}, then such a phenomenon may be ubiquitous in early-Y dwarfs. We find no obvious evidence of the presence of NH$_3$ in the Y-band spectra for any of our late-type dwarfs. 

\subsection{Colors}
We synthesized $Y-$, $J-$, and $H-$band photometry using the MKO-NIR photometric system for each brown dwarf in our sample.  We also synthesized {\it HST} F105W, F125W, and F140W photometry using available system throughput tables\footnote{http://www.stsci.edu/hst/wfc3/ins\_performance/throughputs}.  Uncertainties are computed using a Monte Carlo approach.  For each spectrum, a value is selected from a normal distribution using the flux density at each wavelength as the mean and the flux density uncertainty as the standard deviation.  Synthetic photometry is performed on the resulting spectrum, and the process is repeated 1000 times.  We take the mean photometric value as the true value, and the standard deviation of the distribution as the uncertainty.  The resulting magnitudes and uncertainties are listed in Table 5.  To investigate trends as a function of spectral type, we also synthesize photometry on the entire sample of T dwarfs available from the Spex Prism Spectral Library\footnote{http://pono.ucsd.edu/$\sim$adam/browndwarfs/spexprism/}(as of June, 2014) using the same method described above.  Colors as a function of spectral type are shown in Figure 13. 

Our synthetic YJH photometry generally agrees well with published photometry from \cite{cush11}, \cite{kirk12}, \cite{mor12}, \cite{dup13}, \cite{leg13}, \cite{beich14}, \cite{cush14a}, \cite{kirk14}, and \cite{leg14b}, with only a few exceptions (56 out of 62 (87\%) measurements have differences that are within 3$\sigma$ of zero).  Our measured H magnitude for WISE 0410$+$1502 and J magnitude for WISE 1405$+$5534 differ significantly from the measured values in \cite{cush11} and \cite{kirk12}, though they do agree with the values quoted in \cite{leg13}.  Our J- and H-band magnitudes for WISE 1541$-$2250 disagree with the values given in \cite{mor12} and \cite{leg14b}, respectively.  Using our measured magnitudes, WISE 1541$-$2250 does not stand out prominently from the other Y1 dwarfs in color-color and color versus spectral type diagrams.  Our measured J magnitude for WISE 1738$+$2732 disagrees with the value quoted in \cite{leg13}, though does agree with the measurements from \cite{cush11} and \cite{kirk12}.  Lastly, our measured J and H magnitudes for WISE 2056$+$1459 agree well with the measured values from \cite{cush11} and \cite{kirk12}, though do not agree with the values from \cite{leg13}.  While we cannot rule out extreme variability as the source of magnitude differences, it is more likely that some measurements are erroneous.  Since our photometry comes from our slitless grism spectroscopy, one potential source of error in our measurements is contamination from nearby bright objects or uneven background fluctuations, though our reduction process (Section 3.2) should minimize such effects.    

Previous investigations of color/spectral type relations show distinct changes at the T/Y boundary (\citealt{lod13}, \citealt{leg13}, \citealt{leg14b}).  We see similar changes in Figure 13, namely, the $Y-J$ (and $F105W-F125W$) color plummets for spectral types later than T8, and the $J-H$ color plateaus at $\sim$T7 and slowly turns redder for the Y dwarfs.  As noted by \cite{leg10}, the blueward trend in the $Y-J$ colors of late type dwarfs may be due to K I condensing into KCl as temperatures decrease.  This condensing weakens the broad 0.77 $\mu$m K I doublet, resulting in a brighter Y-band.  A similar trend is seen the the $z-J$ colors of Y dwarfs in \cite{lod13}.  The redward trend of the $J-H$ color for the latest Y dwarfs occurs because the J-band peak for these objects continues to narrow, while the H-band peak does not (see Figure 11).  

Figure 14 shows the $Y-J$ (when available) versus $J-H$ and $J-H$ versus $J-W2$ colors for our complete sample of brown dwarfs along with synthesized colors from the model spectra of \cite{sau12}, \cite{mor12}, and \cite{mor14}.  While neither the cloudy or cloud-free models reproduce the trends seen for our measured dwarfs, the cloudy models do a better job approximating the colors of our late-type dwarf sample than the cloud-free models. \cite{mor12} show that the inclusion of clouds is an efficient way to redden the $J-H$ colors of late type brown dwarfs.  Figure 14 also shows that models with lower surface gravities and lower $f_{sed}$ values do not fit the observed color trends as well.         
  
\begin{figure*}
\plotone{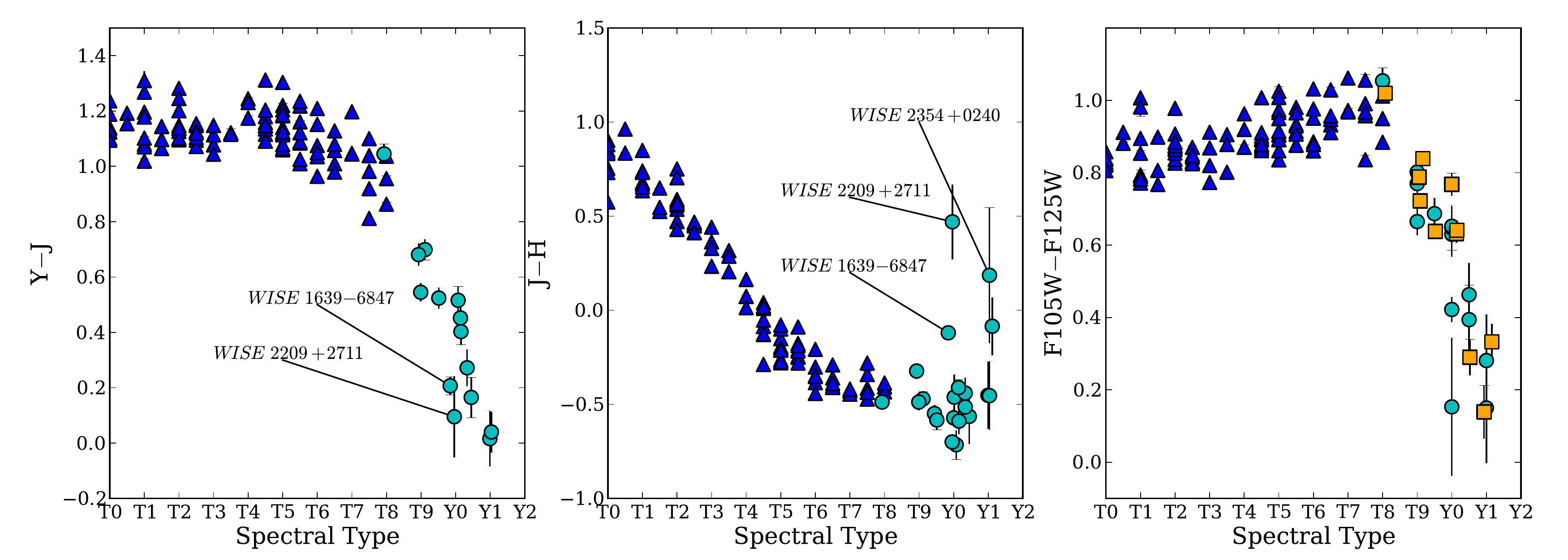}
\caption{{\it Left:} $Y-J$ color as a function of spectral type for T and Y dwarfs. T dwarfs from the SpeX Prism Spectral Library are represented by blue triangles, while synthesized colors of brown dwarfs in this study are represented by cyan circles.  Small offsets have been added along the abscissa for differentiation purposes. {\it Center:} $J-H$ color as a function of spectral type for T and Y dwarfs. {\it Right:} {\it HST} colors as a function of spectral type for T and Y dwarfs. Synthetic photometry of brown dwarfs in this study are represented by cyan circles, while colors found using aperture photometry are represented by orange squares.}
\end{figure*}

\begin{figure*}
\plotone{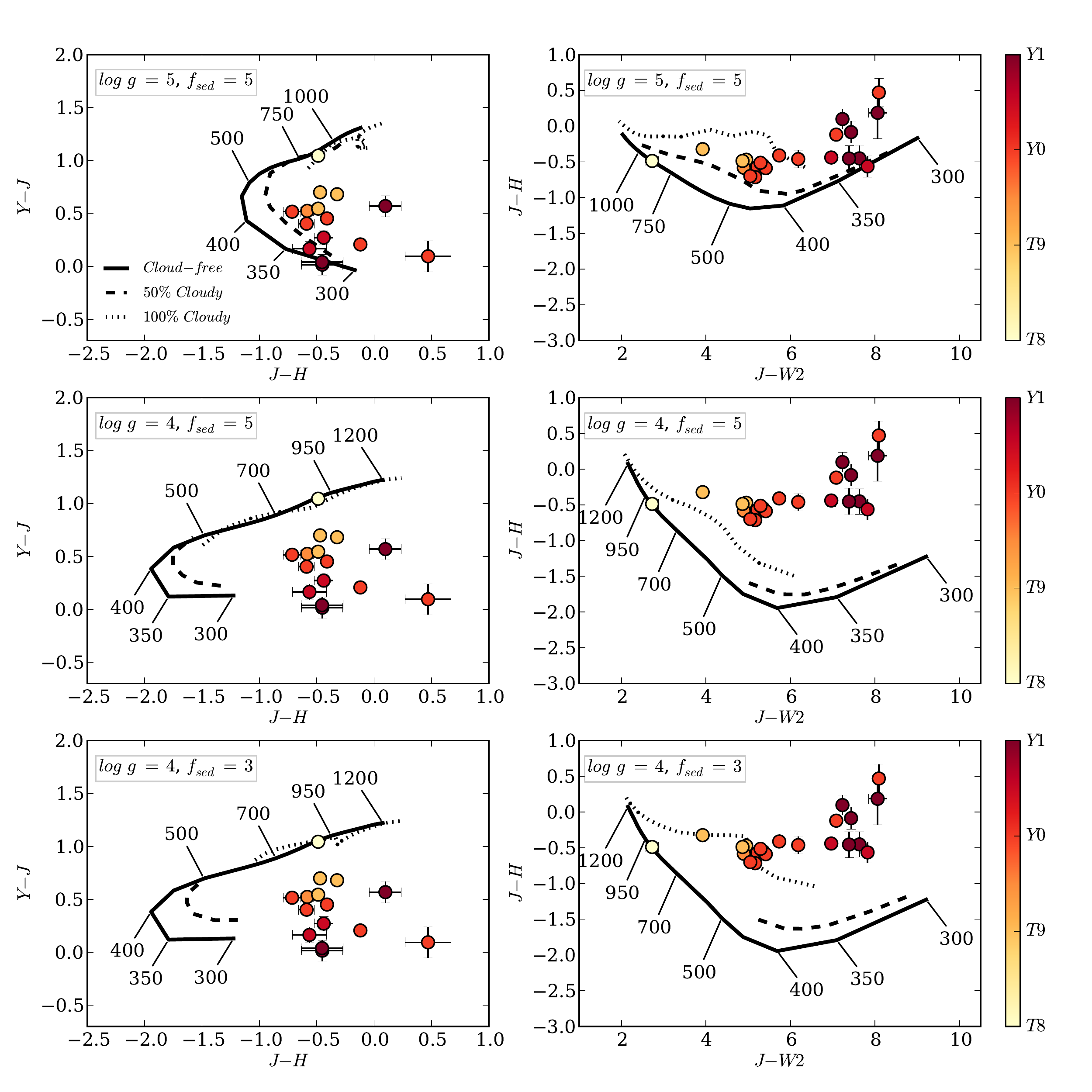}
\caption{{\it Left:}  $J-H$ versus $Y-J$ synthesized colors for brown dwarfs from our sample, cloud-free \cite{sau12} models, and cloudy models from \cite{mor12} and \cite{mor14}.  {\it Right:} $J-W2$ versus $J-H$ for brown dwarfs from our sample with the \cite{sau12}, \cite{mor12}, and \cite{mor14} models.  $f_{sed}$ values apply to the cloudy and partly cloudy models (not the cloud-free models) in all panels.}
\end{figure*}
   
\subsection{Physical Parameters}

The effective temperatures of brown dwarfs are typically estimated in two ways.  If the distance to a brown dwarf is known and a significant fraction of its emergent flux can be measured, then its bolometric luminosity can be computed.  Since brown dwarfs all have similar radii due to the equation of state of their partial degenerate interiors, an effective temperature can then be computed using the Stefan Boltzmann law (e.g. \citealt{gol04}, \citealt{vrba04}, \citealt{dup13}).  Alternatively, model spectra can be compared directly to observed spectra and/or photometry to infer not only the effective temperatures of brown dwarfs, but also their surface gravity and metallicity (e.g.\ \citealt{cush08}, \citealt{witte11}, and \citealt{leg13}).  Because accurate distances to many of the dwarfs in our sample are not known, we have taken the second approach and compared our near-infrared spectra and mid-infrared photometry to atmospheric models in order to estimate their atmospheric parameters.

In particular, we compare are spectra and photometry to several different sets of solar metallicity models, including the cloud-free models from \cite{sau12}, the cloudy models of \cite{mor12}, and the cloudy models of \cite{mor14}.  The latter set is a new suite of models that in addition to including sulfide and chloride condensates, also includes water ice clouds for objects with T$_{eff}$ $\leq$ 450 K.  Some models from \cite{mor14} are also ``partly cloudy'', calculated with 50\% cloudy and 50\% hole surface coverage.  The models are arranged in a grid of T$_{eff}$, log $g$, and $f_{sed}$ values.  The $f_{sed}$ parameter describes the efficiency of cloud sedimentation in the models, where larger $f_{sed}$ values correspond to larger grain sizes, which lead to an increased sedimentation efficiency, and hence thinner clouds.  We adopt the $f_{sed}$ = ``nc'' or ``no clouds'' notation of \cite{cush08} to refer to the cloud-free models.  Table 6 summarizes the T$_{eff}$, log $g$, and $f_{sed}$ combinations that were used for model fitting.

In order to compare the models to our observed spectra, each model spectrum was resampled to be uniform in ln($\lambda$) space, smoothed with a Gaussian kernel to a resolving power of {\it HST} WFC3, and resampled onto the same wavelength grid as the {\it HST} WFC3 spectrum.  Synthetic flux density values in the [3.6] and [4.5] {\it Spitzer} bands were computed from the models as described in Cushing et al. (2006).  Following \cite{cush08}, we evaluate a goodness of fit parameter for each model spectrum, defined as

\begin{equation}
G_{k} = \sum\limits_{i=1}^n \left(\frac{f_i - C_k\mathcal{F}_{k,i}}{\sigma_i}\right)^2
\end{equation}

\noindent where $n$ corresponds to each data pixel, $f_i$ is the flux density of the data, $\mathcal{F}_{k,i}$ is the flux density of the model $k$, and $\sigma_i$ are the errors for each observed flux density.  The $C_k$ parameter is a multiplicative constant given by

\begin{equation}
C_{k} = \frac{\sum f_i\mathcal{F}_{k,i}/\sigma_i^2}{\sum \mathcal{F}_{k,i}/\sigma_i^2}.
\end{equation}

\noindent $C_k$ is equivalent to $(R^2/d^2)$, where $R$ is the stellar radius and $d$ is the distance.  For each model we compute a $C_{k}$ value and corresponding $G_{k}$ value and identify the best fitting model as that model with the lowest overall $G_{k}$ value.  Because in most instances the difference between the best fitting models for each dwarf are marginal, we proved a range of best fit model parameters from the three best fitting models for each brown dwarf in Table 7.  The ($R^2$/$d^2$) and $\chi^2$ values in Table 7 refer to the model with the best fit.

The derived effective temperatures are generally correlated with spectral type in that the effective temperatures decrease with increasing spectral type: WISE 0325$-$5044 (the lone T8) has an effective temperature range of 550$-$600 K, the T9 dwarfs have temperatures between 500 and 600 K, the T9.5 dwarfs have temperatures between 450 and 500 K, the Y0 dwarfs have temperatures between 400 and 450 K, the Y1 dwarfs have temperatures between 300 and 400 K.  Exceptions include the Y0: dwarf WISE 2209$+$2711 (T$_{eff}$ = 500$-$550 K) and the Y1 dwarf WISE 0535$-$7500 which is located in an extremely crowded field and thus has contamination issues (see Section 5.1).

A comparison with the effective temperatures derived from bolometric luminosities in \cite{dup13} find some agreement, though there are several discrepancies.  The main discrepancies appear for the objects with very uncertain parallaxes in \cite{dup13} (WISE 0359$-$5401, WISE 0535$-$7500, WISE 1541$-$2250, and WISE 1639$-$6847).  For these objects with uncertain parallaxes, two (WISE 0359$-$5401 and WISE 1639$-$6847) are found to have T$_{eff}$ values much colder in \cite{dup13} than our model-fitting derived values, while the other two (WISE 0535$-$7500 and WISE 1541$-$2250) are found to be much warmer.  It is unclear whether these discrepancies are solely due to parallax uncertainties or other systematic errors (see below).


Although atmospheric models can be computed for any given T$_{eff}$/log $g$ values, not all combinations are physical because theoretical evolutionary models limit the T$_{eff}$/log $g$ values that brown dwarfs can obtain.  We therefore compared our derived T$_{eff}$/log $g$ values to the low-mass, solar metallicity, cloud-free evolutionary models of \cite{sau08} in order to confirm that our best fit models are all physical (Figure 15). We note that the atmospheric models used in our analysis were not used as boundary conditions for the evolutionary models and thus our comparisons are not technically self consistent.  Inspection of Figure 15 suggests that several of the derived surface gravities are not physical (those objects with log $g$ values of 5 and $T_{eff}$ estimates $\leq$ 400 K).  Evolutionary models can also be used to estimate the ages of brown dwarfs.  Most of our sources have ages of several Gyr, as expected for objects near to the Sun, though some objects were found to have younger (a few hundred Myr) or older ($>$ 5 Gyr) ages than expected.  Indeed, a trend is apparent whereby the earlier spectral types in our sample (T8 and T9) are preferentially fit with lower log $g$ values and hence younger ages, while the latest spectral types (Y0.5 and Y1) are preferentially fit with larger log $g$ values (some of which are unphysical).  Although it is possible this trend is real, it is more likely a result of systematic errors in the model fits.  Fits with unphysical gravities and are indicated as such in the age column of Table 7.

\begin{figure*}
\plotone{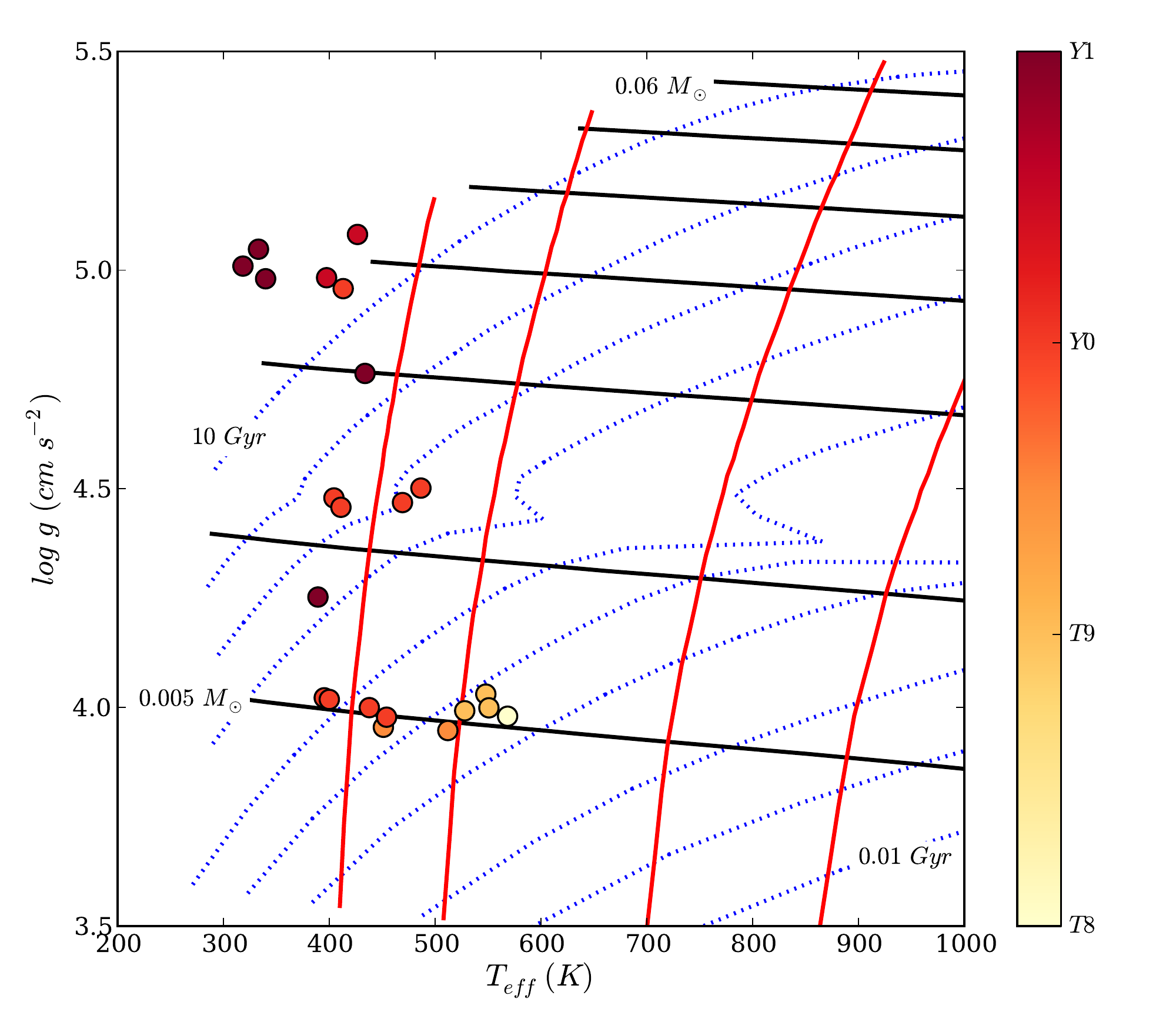}
\caption{The best fitting model parameters from Table 7 compared with the cloud-free evolutionary models from \cite{sau08}.  Lines of constant mass are plotted in solid black for 0.005, 0.01, 0.02, 0.03, 0.04, 0.05, 0.06 $M_\sun$ from bottom to top.  Lines of constant luminosity are plotted in solid red for -6.4, -6.0, -5.4, and -5.0 log(L/$L_\sun$) from left to right.  Isochrones are plotted as dotted blue lines for 10, 4, 2, 1, 0.4, 0.2, 0.1, 0.04, 0.02, and 0.01 Gyr from left to right.  Small offsets have been added along the abscissa and ordinate for differentiation purposes.}
\end{figure*}

While the derived temperatures are generally consistent with previous work, the fits to the data are, in many cases, rather poor.  The discrepancies between the data and models generally fall into two categories.  Because the {\it HST} spectra contain over two hundred data points, the two IRAC points carry little weight in the fitting process and thus the models ``match'' the near-infrared data reasonably well but often miss the {\it Spitzer}/IRAC points badly.  Figure 16 shows the best fits to the T8 to Y1 spectral sequence from Figure 12.  In some instances (e.g. WISE 0325$-$5044, WISE 1541$-$2250), the {\it Spitzer} model fluxes can be off by factors of 0.2 to 4, respectively.  Since much of the energy of cold brown dwarfs emerges at 5 $\mu$m, the bolometric luminosity of the best-fitting models disagree with the observations by similar factors.  A factor of 2 to 5 in $L_{bol}$ can result an effective temperature difference of up to 200 K (see Figure 15).

\begin{figure*}
\plotone{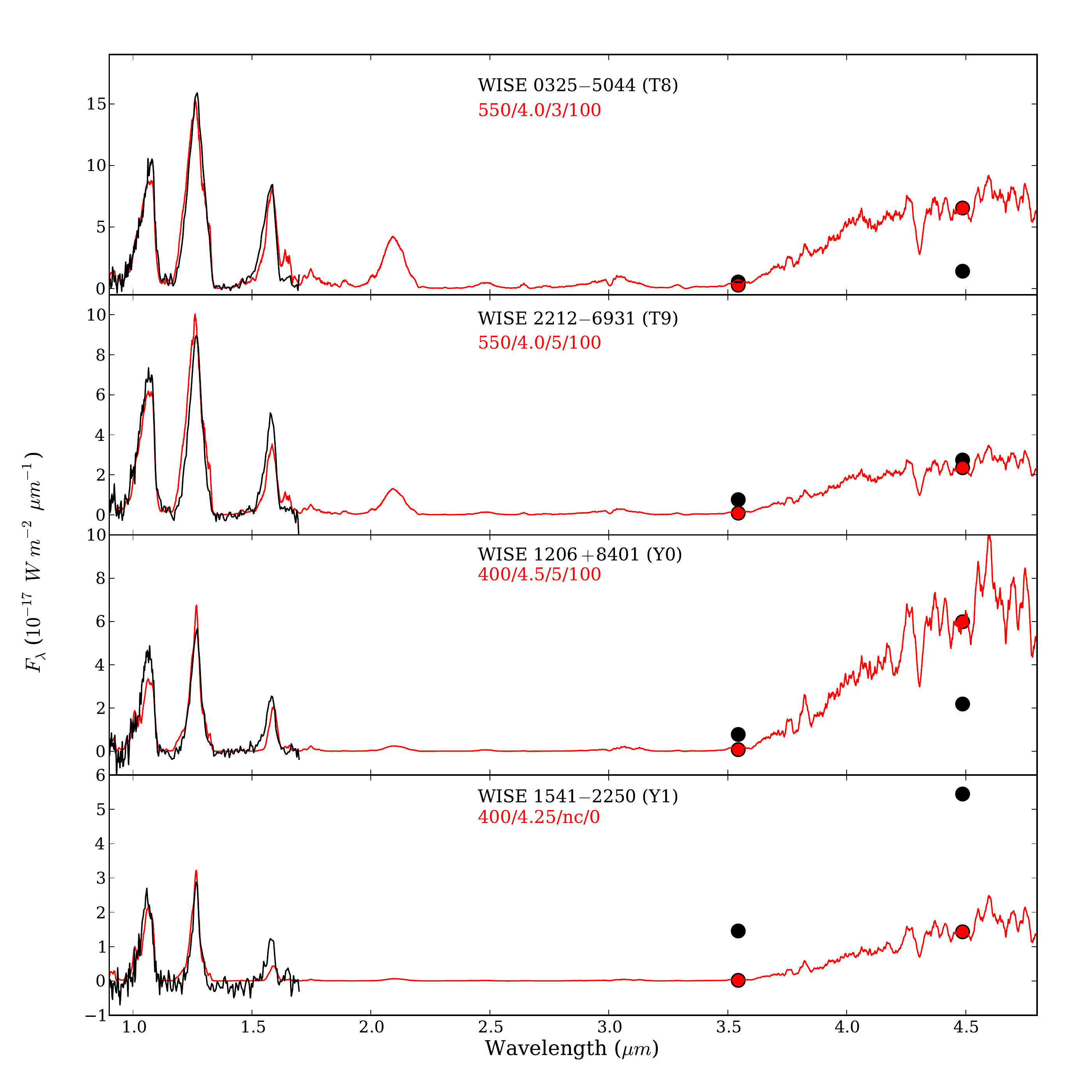}
\caption{The T8 to Y1 spectral sequence (black) with best fit models (red).  The best fitting model parameters (T$_{eff}$ in K, log $g$ in cm s$^{-1}$, $f_{sed}$, and the percentage of cloud cover) are given under each object name.  The black symbols represent the {\it Spitzer} IRAC photometry from Table 2, while the red circles indicate the synthetic photometry of the model in the {\it ch1} and {\it ch2} bandpasses. Error bars for the IRAC photometry are smaller than the symbol size.}
\end{figure*}

Closer inspection also reveals large differences between the models and observations in the shapes of the near-infrared bands.  In particular, the heights of the model $H$-band peaks are significantly different than the observations, with the observed H-band peak consistently being higher then that of the best-fitting model.  The height, width, and position of the J-band peak is generally well matched to the models.  For the two latest spectral types, the J-band peak of the observations does not reach the same heights as those from the best-fitting models.  The Y-band peaks for the two T dwarfs match well with the models, while the fits to the Y-band peaks of the Y dwarfs do not.  The main discrepancy in the Y-band spectra of the Y dwarfs is the absence in the observations of the NH$_3$ absorption feature expected to occur for temperatures $<$500 K.  The problems with these particular model fits are nicely summarized by the color-color diagrams shown in Figure 14.

Some of the issues in the near-infrared are likely a result of the fact that the models do not account for non-equilibrium NH$_3$ chemistry due to vertical mixing in the atmosphere (see Section 5.2).  Figure 17 shows the near-infrared spectrum of WISE 1541$-$2250 (Y1) along with the cross section spectrum of NH$_3$.  Ammonia absorption shapes the entire near-infrared spectrum -- from the blue wing of the $H$-band peak, to the width of the $J$-band peak, to the shape of the $Y$-band peak.  Morley et al. (2014) show that when all else is equal, the decrease in the NH$_3$ abundance broadens the $J$- and $H$-band peaks, and removes the strong NH$_3$ absorption feature in the $Y$ band.  The inclusion of non-equilibrium chemistry, which would result in significant changes to the model spectra would no doubt result in different effective temperature and surface gravity estimates.  Based on the $H$ band alone, the change would be 300 K at a fixed gravity. Given the significant mismatches identified between these model spectra and data, we conclude that we cannot reliably derive reliable atmospheric parameters by fitting these model spectra to the observed spectra and photometry of our sample.  
  
\begin{figure}
\plotone{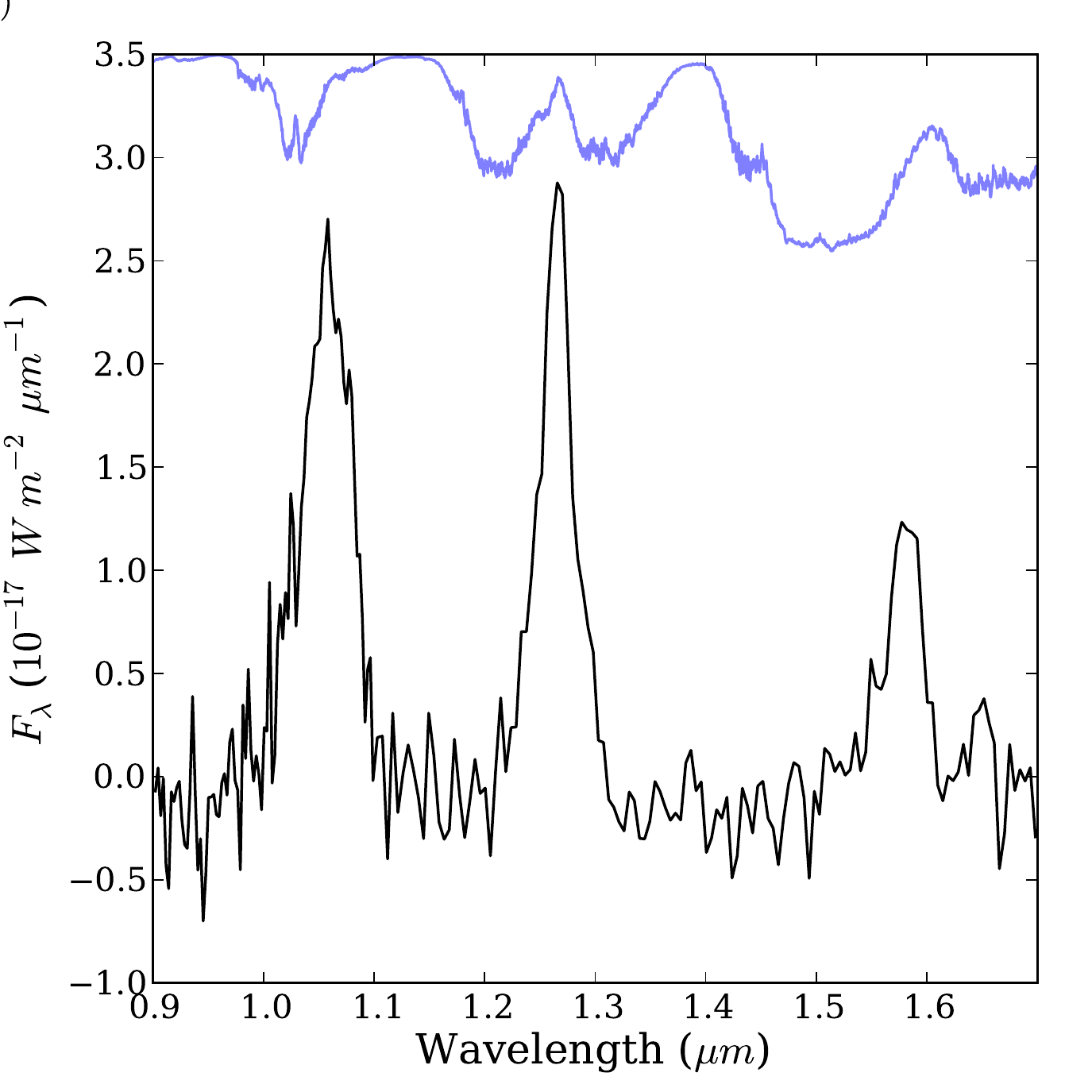}
\caption{The near-infrared spectrum of WISE 1541$-$2250.  The cross section spectrum of NH$_3$ at T = 600 K and P = 1 bar (Richard Freedman, priv.\ comm.), indicating the location of prominent NH$_3$ absorption bands, is shown in blue.  The NH$_3$ cross section spectrum is normalized at 1.033 $\mu$m, inverted, and offset for clarity.}
\end{figure}

\cite{leg14b} state that the $Y$, $H$, and {\it ch1} model fluxes are too low by about a factor of two compared to photometric measurements for late-type brown dwarfs.  By extension, this suggests that the $J$ and $W2$ magnitudes are not affected and thus the $J-W2$ color may be able to provide a reasonable effective temperature estimate when compared with colors from model spectra.  Figure 18 shows the $J-W2$ color as a function of effective temperature for three different models - a cloud-free model from \cite{sau12} (log $g$ = 5), a 50\% cloud coverage model from \cite{mor14} (log $g$ = 5, $f_{sed}$ = 5, with $H_2O$), and a 100\% cloud coverage model from \cite{mor12} (log $g$ = 5, $f_{sed}$ = 3, without $H_2O$).  Because our grid of models is, as of yet, incomplete (see Table 6), we show the only 50\% cloud coverage model (with $H_2O$) for which the entire 300$-$1000 K temperature range is available and the 100\% cloud coverage model does not extend to temperatures $<$ 400 K.  Even when limited to these three models, Figure 18 shows that effective temperature estimates for $J-W2$ values less than $\sim$7 typically have ranges greater than 100 K.  For $J-W2$ values greater than $\sim$7, varying percentages of cloud cover result in minimal differences in effective temperature, though the 100\% cloudy model does not extend to this range.  The range of temperatures for a single $J-W2$ value will only increase with the inclusion of additional models with varying log $g$ and $f_{sed}$ values.     

\begin{figure}
\plotone{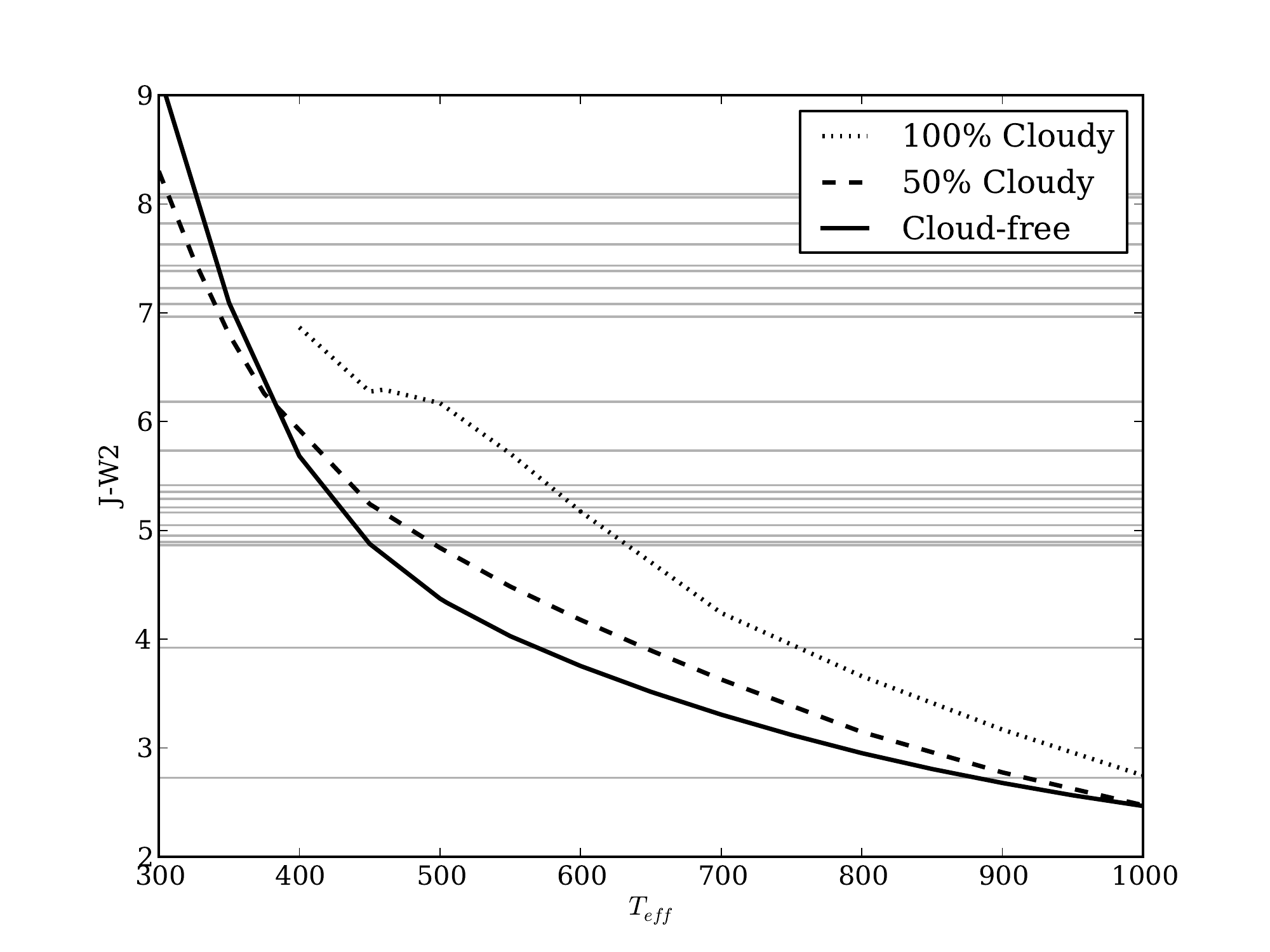}
\caption{The $J-W2$ color as a function of effective temperature for a cloud-free model model from \cite{sau12} (log $g$ = 5), a 50\% cloud coverage model from \cite{mor14} (log $g$ = 5, $f_{sed}$ = 5, with $H_2O$), and a 100\% cloud coverage model from \cite{mor12} (log $g$ = 5, $f_{sed}$ = 3, without $H_2O$).   Each horizontal gray line indicates a single $J-W2$ value for a brown dwarf in our sample.}
\end{figure}

Furthermore, as noted previously, the $J-$band flux will also be affected by the reduced amount of NH$_3$ present in the atmosphere due to vertical mixing (not only $Y$ and $H$).  Figure 12 of \cite{mor14} displays equilibrium and disequilibrium models at 450, 300, and 200 K.  For the 450 K models, including disequilibrium chemistry will change the $Y-$, $J-$, and $H-$ band magnitudes by $\sim$0.2, $\sim$0.3, and $\sim$0.6, respectively, implying that these near-infrared bands are affected by similar amounts when vertical mixing is included.  Because of the large temperature ranges for individual $J-W2$ values, uncertain temperature ranges for $J-W2$ colors due to of our incomplete model grid, and the similar influence of vertical mixing on the $J-$band magnitude when compared to the $Y-$ and $H-$ bands, we conclude that estimating temperatures based solely on model $J-W2$ colors is unsound.  Since our observations are limited to near- and mid-infrared wavelengths and the current models do not include non-equilibrium chemistry, we lack a method to determine accurate effective temperatures for the objects in our sample.     

Given the limitations of these particular models, determining effective temperatures by computing bolometric luminosities may therefore be preferable (e.g.\ \citealt{dup13}).  However, determining luminosities in this way currently relies on models to estimate the flux levels for up to 50\% of the total luminosity \citep{dup13}.  \cite{dup13} note that all current atmospheric models provide similar bolometric corrections in the wavelength ranges where flux has not yet been directly measured. Nevertheless, minimizing the dependence of the derived bolometric luminosities on the models should be a high priority, and therefore determining accurate effective temperatures for the lowest mass brown dwarfs should focus on obtaining flux measurements at wavelengths longer than $\sim$5 $\mu$m.  The near-infrared (NIRSpec and NIRCam) instruments and in particular the mid-infrared instrument (MIRI) aboard the forthcoming {\it James Webb Space Telescope} could prove invaluable in providing the additional flux coverage needed to determine accurate effective temperatures for the lowest mass brown dwarfs.           
                     
\section{Conclusions}
We have presented  {\it HST} near-infrared spectra for a sample of twenty-two brown dwarfs, six of which are new discoveries.  Three of the new discoveries are classified as having spectral type Y, bringing the total number of spectroscopically confirmed Y dwarfs to twenty-one.  Theoretical spectra for the lowest temperature brown dwarfs do not yet accurately reproduce the spectra from these observations.  While disagreements between models and observations have been known to exist for these objects, the inclusion of {\it HST} G102 spectra has made discrepancies even more apparent.  The primary disagreements occur in regions where models predict detectable NH$_3$.  These spectra, as well as those in \cite{leg13}, \cite{leg14}, and \cite{leg14b}, show that the strong NH$_3$ absorption features expected to occur in the Y- and H-band spectra of Y dwarfs are not present.  While the non-equilibrium chemistry for T dwarfs reproduce observations quite well \citep{hub07}, the latest non-equilibrium models do not yet agree with the observations \citep{mor14}.   That the models diverge from accurately describing the spectra of objects at such low temperatures begs for new models that include improved non-equilibrium chemistry, or the identification of another method for reducing NH$_3$ abundances.  
 
\acknowledgments

We wish to thank the referee, Sandy Leggett, for numerous suggestions that improved this manuscript.  We wish to thank Caroline Morley and Didier Saumon for useful discussions regarding low temperature models.  This publication makes use of data products from the {\it Wide-field Infrared Survey Explorer}, which is a joint project of the University of California, Los Angeles, and the Jet Propulsion Laboratory/California Institute of Technology, and NEOWISE, which is a project of the Jet Propulsion Laboratory/California Institute of Technology. WISE and NEOWISE are funded by the National Aeronautics and Space Administration.  This research has benefitted from the M, L, T, and Y dwarf compendium housed at dwarfarchives.org.  This research has benefitted from the SpeX Prism Spectral Libraries, maintained by Adam Burgasser at http://pono.ucsd.edu/$\sim$adam/browndwarfs/spexprism. HST acknowledgement needed.  We thank the STSCI help desk for useful discussions and resolution suggestions regarding WFC3 IR photometry.  The authors wish to thank Caroline Morley for providing spectroscopic models via the webpage http://www.ucolick.org/$\sim$cmorley/cmorley/Models.html.

\clearpage 
\begin{center}

\begin{deluxetable}{lcccccl}
\tablecaption{AllWISE Photometry}
\tabletypesize{\scriptsize}
\tablewidth{0pt}
\tablehead{
\colhead{AllWISE Name} & \colhead{Spec.} & W1 & W2 & W3 & W1$-$W2\\
& Type & (mag) & (mag) & (mag) & (mag)}
\startdata
\cutinhead{New Discoveries}
WISEA J032504.52$-$504403.0 & T8 & 18.430 $\pm$ 0.258 & 16.209 $\pm$ 0.145 & $>$12.918 & 2.22 $\pm$ 0.30\\
WISEA J040443.50$-$642030.0 & T9 & 18.442 $\pm$ 0.178 & 15.726 $\pm$ 0.063 & $>$12.977 & 2.72 $\pm$ 0.19\\
WISEA J082507.37$+$280548.2 & Y0.5 & $>$18.444 & 14.578 $\pm$ 0.060 & $>$11.660 & $>$3.87\\
WISEA J120604.25$+$840110.5 & Y0 & $>$18.734 & 15.058 $\pm$ 0.054 & $>$12.536 & $>$3.68\\
WISEA J221216.27$-$693121.6 & T9 & 17.259 $\pm$ 0.122 & 14.873 $\pm$ 0.061 & $>$12.621 & 2.39 $\pm$ 0.14\\
WISEA J235402.79$+$024014.1 & Y1 & $>$18.263 & 15.007 $\pm$ 0.085 & $>$12.278 & $>$3.26\\
\cutinhead{Other Brown Dwarfs in this Study}
WISEA J033515.07$+$431044.7 & T9 & $>$18.652 & 14.515 $\pm$ 0.055 & $>$11.901 & $>$4.14\\
WISEA J035000.31$-$565830.5 & Y1 & $>$18.699 & 14.745 $\pm$ 0.044 & 12.325 $\pm$ 0.282 & $>$3.95\\
WISEA J035934.07$-$540154.8 & Y0 & $>$19.031 & 15.384 $\pm$ 0.054 & $>$12.877 & $>$3.65\\
WISEA J041022.75$+$150247.9 & Y0 & $>$18.170 & 14.113 $\pm$ 0.047 & 12.314 $\pm$ 0.500 & $>$4.06\\
WISEA J053516.87$-$750024.6 & $\geq$Y1 & 17.940 $\pm$ 0.143 & 14.904 $\pm$ 0.047 & $>$12.349 & 3.04 $\pm$ 0.15\\
WISEA J064723.24$-$623235.4 & Y1 & $>$19.539 & 15.224 $\pm$ 0.051 & $>$12.961 & $>$4.32\\
WISEA J073444.03$-$715743.8 & Y0 & 18.749 $\pm$ 0.281 & 15.189 $\pm$ 0.050 & $>$12.959 & 3.56 $\pm$ 0.29\\
WISEA J094306.00$+$360723.3 & T9.5 & 18.176 $\pm$ 0.297 & 14.413 $\pm$ 0.048 & 12.289 $\pm$ 0.394 & 3.76 $\pm$ 0.30\\
WISEA J140518.32$+$553421.3 & Y0.5 & 18.765 $\pm$ 0.396 & 14.097 $\pm$ 0.037 & 12.204 $\pm$ 0.263 & 4.67 $\pm$ 0.40\\
WISE J154151.65$-$225024.9\tablenotemark{a} & Y1 & 16.736 $\pm$ 0.165\tablenotemark{b} & 14.246 $\pm$ 0.063 & $>$12.312 & 2.49 $\pm$ 0.18\tablenotemark{b}\\
WISEA J154214.00$+$223005.2 & T9.5 & 18.846 $\pm$ 0.425 & 15.043 $\pm$ 0.061 & $>$13.014 & 3.80 $\pm$ 0.43\\
WISEA J163940.84$-$684739.4 & Y0pec & 17.266 $\pm$ 0.187 & 13.544 $\pm$ 0.059 & $>$11.755 & 3.72 $\pm$ 0.20\\
WISEA J173835.52$+$273258.8 & Y0 & 17.710 $\pm$ 0.157 & 14.497 $\pm$ 0.043 & 12.448 $\pm$ 0.399 & 3.21 $\pm$ 0.16\\
WISEA J205628.88$+$145953.6 & Y0 & 16.480 $\pm$ 0.075 & 13.839 $\pm$ 0.037 & 11.731 $\pm$ 0.249 & 2.64 $\pm$ 0.08\\
WISEA J220905.75$+$271143.6 & Y0: & $>$18.831 & 14.770 $\pm$ 0.055 & 12.455 $\pm$ 0.387 & $>$4.06\\
WISEA J222055.34$-$362817.5 & Y0 & $>$18.772 & 14.714 $\pm$ 0.056 & $>$12.292 & $>$4.06\\  
\cutinhead{Interlopers}
WISEA J013810.99$+$201657.5 & \dots & 18.120 $\pm$ 0.273 & 15.221 $\pm$ 0.087 & 11.670 $\pm$ 0.199 & 2.90 $\pm$ 0.29\\
WISEA J065954.18$-$585559.6 & \dots & 17.897 $\pm$ 0.133 & 15.439 $\pm$ 0.059 & 12.955 $\pm$ 0.371 & 2.46 $\pm$ 0.15  
\enddata
\tablenotetext{a}{WISE 1541$-$2250 is not in the AllWISE catalog.}
\tablenotetext{b}{Details of the likely erroneous W1 measurement of WISE 1541$-$2250 are discussed in \cite{kirk12}.}
\end{deluxetable}

\clearpage

\begin{deluxetable}{lccccccc}
\tablecaption{HST and Spitzer Aperture Photometry}
\tabletypesize{\scriptsize}
\tablewidth{0pt}
\tablehead{
\colhead{AllWISE Name} & F105W & F125W & F140W & {\it ch1} & {\it ch2} & {\it ch1$-$ch2}\\
& (mag) & (mag) & (mag) & (mag) & (mag) & (mag)}
\startdata
WISEA J032504.52$-$504403.0 & 20.618 $\pm$ 0.009 & 19.598 $\pm$ 0.002 & \dots & 17.746 $\pm$  0.086 & 15.696 $\pm$ 0.025 & 2.050 $\pm$ 0.090 \\
WISEA J033515.07$+$431044.7 & 20.880 $\pm$ 0.015 & 20.092 $\pm$ 0.003 & \dots & 16.612 $\pm$ 0.040 & 14.381 $\pm$ 0.020 & 2.231 $\pm$ 0.045 \\
WISEA J035000.31$-$565830.5 & \dots & \dots & 22.321 $\pm$ 0.047 & 17.936 $\pm$ 0.096 & 14.688 $\pm$ 0.020 & 3.248 $\pm$ 0.098 \\
WISEA J035934.07$-$540154.8 & \dots & \dots & 21.806 $\pm$ 0.039 & 17.553 $\pm$ 0.072 & 15.326 $\pm$ 0.023 & 2.227 $\pm$ 0.075 \\
WISEA J040443.50$-$642030.0 & 21.115 $\pm$ 0.013 & 20.276 $\pm$ 0.003 & \dots & 17.633 $\pm$ 0.082 & 15.418 $\pm$ 0.022 & 2.216 $\pm$ 0.085 \\
WISEA J041022.75$+$150247.9 & \dots & \dots & 19.634 $\pm$ 0.007 & 16.636 $\pm$ 0.042 & 14.166 $\pm$ 0.019 & 2.470 $\pm$ 0.046 \\
WISEA J053516.87$-$750024.6 & 23.140 $\pm$ 0.049 & 22.801 $\pm$ 0.051 & 22.422 $\pm$ 0.050 & 17.753 $\pm$ 0.084 & 15.009 $\pm$ 0.021 & 2.744 $\pm$ 0.087 \\
WISEA J064723.24$-$623235.4 & 23.592 $\pm$ 0.054 & 23.453 $\pm$ 0.050 & \dots & 17.893 $\pm$ 0.092 & 15.070 $\pm$ 0.022 & 2.823 $\pm$ 0.094 \\
WISEA J073444.03$-$715743.8 & 21.732 $\pm$ 0.031 & 20.964 $\pm$ 0.004 & \dots & 17.649 $\pm$ 0.077 & 15.213 $\pm$ 0.022 & 2.436 $\pm$ 0.080 \\
WISEA J082507.37$+$280548.2 & 23.487 $\pm$ 0.040 & 23.197 $\pm$ 0.030 & \dots & 17.624 $\pm$ 0.077 & 14.637 $\pm$ 0.020 & 2.987 $\pm$ 0.079 \\
WISEA J094306.00$+$360723.3 & \dots & \dots & 20.037 $\pm$ 0.007 & 16.746 $\pm$ 0.043 & 14.284 $\pm$ 0.019 & 2.461 $\pm$ 0.047 \\
WISEA J120604.25$+$840110.5 & 21.694 $\pm$ 0.025 & 21.062 $\pm$ 0.005 & \dots & 17.339 $\pm$ 0.061 & 15.220 $\pm$ 0.022 & 2.119 $\pm$ 0.065 \\
WISEA J140518.32$+$553421.3 & 21.939 $\pm$ 0.024 & \dots & 21.271 $\pm$ 0.018 & 16.876 $\pm$ 0.046 & 14.058 $\pm$ 0.019 & 2.818 $\pm$ 0.050 \\
WISE J154151.65$-$225024.9 & 22.204 $\pm$ 0.044 & 21.871 $\pm$ 0.023 & \dots & 16.658 $\pm$ 0.042 & 14.228 $\pm$ 0.019 & 2.430 $\pm$ 0.046 \\
WISEA J154214.00$+$223005.2 & 21.268 $\pm$  0.020 & 20.630 $\pm$  0.003 & 20.240 $\pm$  0.010 & 17.257 $\pm$ 0.059 & 15.057 $\pm$ 0.022 & 2.200 $\pm$ 0.062 \\
WISEA J163940.84$-$684739.4 & 21.337 $\pm$ 0.026 & 21.151 $\pm$ 0.012 & \dots & \dots & 13.537 $\pm$ 0.017 & \dots \\
WISEA J173835.52$+$273258.8 & \dots & \dots & 19.883 $\pm$ 0.008 & 17.093 $\pm$ 0.053 & 14.473 $\pm$ 0.019 & 2.620 $\pm$ 0.056 \\
WISEA J205628.88$+$145953.6 & \dots & \dots & 19.524 $\pm$ 0.007 & 16.031 $\pm$ 0.030 & 13.923 $\pm$ 0.018 & 2.108 $\pm$ 0.035 \\
WISEA J220905.75$+$271143.6 & 23.842 $\pm$ 0.057 & \dots & 23.167 $\pm$ 0.149 & 17.815 $\pm$ 0.087 & 14.739 $\pm$ 0.020 & 3.076 $\pm$ 0.090 \\
WISEA J221216.27$-$693121.6 & 21.069 $\pm$ 0.012 & 20.347 $\pm$ 0.003 & \dots & 17.364 $\pm$ 0.063 & 14.973 $\pm$ 0.021 & 2.391 $\pm$ 0.066 \\
WISEA J222055.34$-$362817.5 & 21.638 $\pm$ 0.027 & 20.997 $\pm$ 0.005 & \dots & 17.200 $\pm$ 0.057 & 14.736 $\pm$ 0.021 & 2.464 $\pm$ 0.061 \\
WISEA J235402.79$+$024014.1 & \dots & 23.368 $\pm$ 0.094 & \dots & 18.105 $\pm$ 0.109 & 15.013 $\pm$ 0.022 & 3.091 $\pm$ 0.111
\enddata

\end{deluxetable}

\clearpage

\begin{deluxetable}{lcccc}
\tablecaption{HST/WFC3 Spectroscopy Log}
\tabletypesize{\scriptsize}
\tablewidth{0pt}
\tablehead{
\colhead{AllWISE Name} & \colhead{Date} & mode & Total Int.\ Time & \# of Exp. \\
& \colhead{(UT)} & & (s) & }
\startdata
WISEA J032504.52$-$504403.0 & 2013 July 30 & G102 & 1612 & 4\\
 & 2013 Aug 4 & G141 & 1612 & 4\\
WISEA J033515.07$+$431044.7 & 2013 July 12 & G102 & 1812 & 4\\
 & 2013 Aug 30 & G141 & 1812 & 4\\
WISEA J035000.31$-$565830.5\tablenotemark{a} & 2011 Aug 13 & G141 & 2212 & 4\\
WISEA J035934.07$-$540154.8\tablenotemark{a} & 2011 Aug 10 & G141 & 2212 & 4\\
WISEA J040443.50$-$642030.0 & 2013 April 9 & G102 & 1812  & 4\\
 & 2013 April 9 & G141 & 1812 & 4 \\
WISEA J041022.75$+$150247.9 & 2012 Sep 1 & G141 & 2012 & 4\\
WISEA J053516.87$-$750024.6\tablenotemark{a} & 2011 Sep 27 & G141 & 2212 & 4\\
 & 2012 Sep 17 & G141 & 2212 & 4\\
 & 2013 Sep 26 & G102 & 7618 & 6\\
 & 2013 Sep 27 & G141 & 7615 & 6\\
 & 2013 Sep 27 & G102 & 7618 & 6\\
 & 2013 Dec 4 & G102 & 7618 & 6\\
WISEA J064723.24$-$623235.4\tablenotemark{b} & 2013 May 13 & G141 & 7218 & 6\\
 & 2013 May 13 & G102 & 7036 & 12\\
 & 2013 May 15 & G102 & 7036 & 12\\
 & 2013 Nov 14 & G102 & 7036 & 12\\
 & 2013 Dec 28 & G102 & 7036 & 12\\
 & 2013 Dec 29 & G141 & 7218 & 6\\
 & 2013 Dec 30 & G102 & 7036 & 12\\
WISEA J073444.03$-$715743.8 & 2013 May 18 & G102 & 1812 & 4\\
 & 2013 May 20 & G141 & 1812 & 4\\
WISEA J082507.37$+$280548.2 & 2014 Jan 16 & G102 & 7018 & 6\\
 & 2014 Jan 17 & G141 & 7218 & 3\\
 & 2014 Jan 18 & G102 & 7018 & 6\\
 & 2014 Jan 19 & G102 & 7018 & 6\\
WISEA J094306.00$+$360723.3 & 2013 Feb 20 & G141 & 2012 & 4\\
WISEA J120604.25$+$840110.5 & 2013 July 15 & G102 & 1812 & 4\\
 & 2013 July 15 & G141 & 1812  & 4\\
WISEA J140518.32$+$553421.3\tablenotemark{c} & 2011 Mar 14 & G141 & 2212 & 4\\
 & 2013 Apr 18 & G102 & 1812 & 4\\
WISE J154151.65$-$225024.9 & 2013 May 9 & G102 & 4612 & 4\\
 & 2013 May 9 & G141 & 1812 & 4 \\
WISEA J154214.00$+$223005.2 & 2012 Mar 4 & G141 & 2012 & 4 \\
 & 2013 June 5 & G102 & 1812 & 4 \\
 & 2013 June 7 & G141 & 1812 & 4 \\
WISEA J163940.84$-$684739.4 & 2013 Oct 26 & G102 & 7518 & 6 \\
 & 2013 Oct 27 & G102 & 7518 & 6 \\
 & 2013 Oct 29 & G141 & 10024 & 4\\
WISEA J173835.52$+$273258.8\tablenotemark{c} & 2011 May 12 & G141 & 2012 & 4\\
WISEA J205628.88$+$145953.6\tablenotemark{a} & 2011 Sept 4 & G141 & 2012 & 4\\
WISEA J220905.75$+$271143.6\tablenotemark{d} & 2012 Sept 15 & G141 & 2012 & 4 \\
 & 2013 Apr 28 & G102 & 7218 & 6\\
 & 2013 June 6 & G102 & 7218 & 6\\
 & 2013 Sept 20 & G102 & 7218 & 6\\
WISEA J221216.27$-$693121.6 & 2013 Sept 7 & G102 & 4612 & 4\\
 & 2013 Sept 11 & G141 & 1812 & 4 \\
WISEA J222055.34$-$362817.5 & 2013 June 8 & G102 & 1812 & 4 \\
 & 2013 June 20 & G141 & 4412 & 4\\
WISEA J235402.79$+$024014.1 & 2013 Sept 22 & G141 & 3224 & 8
\enddata
\tablenotetext{a}{Published previously in \cite{kirk12}.}
\tablenotetext{b}{Published previously in \cite{kirk13}.}
\tablenotetext{c}{Published previously in \cite{cush11}.}
\tablenotetext{d}{Published previously in \cite{cush14a}.}
\end{deluxetable}

\clearpage

\begin{deluxetable}{lccccccc}
\tablecaption{Distance Estimates}
\tabletypesize{\scriptsize}
\tablewidth{0pt}
\tablehead{
\colhead{AllWISE Name} & \colhead{Spec.} & Dist\tablenotemark{a} & Dist\tablenotemark{b} & \\
& Type & (pc) & (pc) }
\startdata
WISEA J032504.52$-$504403.0 & T8 & 36.4$\pm$2.4 & 36.0$\pm$2.4\\
WISEA J040443.50$-$642030.0 & T9 & 24.8$\pm$0.7 & 24.5$\pm$0.7\\
WISEA J082507.37$+$280548.2 & Y0.5 & 10.0$\pm$0.3 & 10.9$\pm$0.3\\
WISEA J120604.25$+$840110.5 & Y0 & 14.5$\pm$0.4 & 15.0$\pm$0.4\\
WISEA J221216.27$-$693121.6 & T9 & 16.8$\pm$0.5 & 16.6$\pm$0.5\\
WISEA J235402.79$+$024014.1 & Y1 & 10.1$\pm$0.4 & 11.9$\pm$0.5
\enddata
\tablenotetext{a}{Using W2 relation of \cite{kirk12}.}
\tablenotetext{b}{Using W2 relation of \cite{dup12}.}
\end{deluxetable}

\begin{deluxetable}{lccccccc}
\tablecaption{Synthetic Photometry}
\tabletypesize{\scriptsize}
\tablewidth{0pt}
\tablehead{
\colhead{AllWISE Name} & \colhead{Spec.} & Y$_{MKO}$ & J$_{MKO}$ & H$_{MKO}$ & F105W & F125W & F140W\\
& Type & (mag) & (mag) & (mag) & (mag) & (mag) & (mag)}
\startdata
WISEA J032504.52$-$504403.0 & T8 & 19.980 $\pm$ 0.027 & 18.935 $\pm$ 0.024 & 19.423 $\pm$ 0.027 & 20.601 $\pm$ 0.028 & 19.547 $\pm$ 0.023 & 19.223 $\pm$ 0.022\\
WISEA J033515.07$+$431044.7 & T9 & 20.166 $\pm$ 0.029 & 19.467 $\pm$ 0.023 & 19.938 $\pm$ 0.031 & 20.939 $\pm$ 0.033 & 20.137 $\pm$ 0.025 & 19.785 $\pm$ 0.023\\
WISEA J035000.31$-$565830.5\tablenotemark{a} & Y1 & \dots & 22.178 $\pm$ 0.073 & 22.263 $\pm$ 0.135 & \dots & 22.951 $\pm$ 0.114 & 22.431 $\pm$ 0.073\\
WISEA J035934.07$-$540154.8\tablenotemark{a} & Y0 & \dots & 21.566 $\pm$ 0.046 & 22.028 $\pm$ 0.112 & \dots & 22.258 $\pm$ 0.062 & 21.789 $\pm$ 0.045\\
WISEA J040443.50$-$642030.0 & T9 & 20.328 $\pm$ 0.032 & 19.647 $\pm$ 0.025 & 19.970 $\pm$ 0.033 & 21.063 $\pm$ 0.037 & 20.293 $\pm$ 0.028 & 19.893 $\pm$ 0.024\\
WISEA J041022.75$+$150247.9\tablenotemark{a} & Y0 & \dots & 19.325 $\pm$ 0.024 & 19.897 $\pm$ 0.038 & \dots & 19.997 $\pm$ 0.025 & 19.643 $\pm$ 0.024\\
WISEA J053516.87$-$750024.6\tablenotemark{b} & $\geq$Y1 & 22.701 $\pm$ 0.070 & 22.132 $\pm$ 0.071 & \dots & 23.581 $\pm$ 0.138 & 22.876 $\pm$ 0.102 & \dots\\
WISEA J064723.24$-$623235.4 & Y1 & 22.870 $\pm$ 0.076 & 22.854 $\pm$ 0.066 & 23.306 $\pm$ 0.166 & 23.833 $\pm$ 0.117 & 23.683 $\pm$ 0.098 & 23.204 $\pm$ 0.066\\
WISEA J073444.03$-$715743.8 & Y0 & 20.870 $\pm$ 0.041 & 20.354 $\pm$ 0.029 & 21.069 $\pm$ 0.071 & 21.675 $\pm$ 0.051 & 21.045 $\pm$ 0.035 & 20.726 $\pm$ 0.030\\
WISEA J082507.37$+$280548.2 & Y0.5 & 22.566 $\pm$ 0.053 & 22.401 $\pm$ 0.050 & 22.965 $\pm$ 0.139 & 23.409 $\pm$ 0.073 & 23.015 $\pm$ 0.062 & 22.731 $\pm$ 0.051\\
WISEA J094306.00$+$360723.3\tablenotemark{a} & T9.5 & \dots & 19.766 $\pm$ 0.025 & 20.315 $\pm$ 0.038 & \dots & 20.444 $\pm$ 0.027 & 20.092 $\pm$ 0.025\\
WISEA J120604.25$+$840110.5 & Y0 & 20.875 $\pm$ 0.036 & 20.472 $\pm$ 0.030 & 21.061 $\pm$ 0.062 & 21.819 $\pm$ 0.050 & 21.171 $\pm$ 0.036 & 20.798 $\pm$ 0.029\\
WISEA J140518.32$+$553421.3 & Y0.5 & 21.333 $\pm$ 0.057 & 21.061 $\pm$ 0.035 & 21.501 $\pm$ 0.073 & 22.193 $\pm$ 0.076 & 21.730 $\pm$ 0.044 & 21.375 $\pm$ 0.035\\
WISE J154151.65$-$225024.9 & Y1 & 21.671 $\pm$ 0.037 & 21.631 $\pm$ 0.064 & 22.085 $\pm$ 0.170 & 22.724 $\pm$ 0.083 & 22.443 $\pm$ 0.097 & 22.138 $\pm$ 0.078\\
WISEA J154214.00$+$223005.2 & T9.5 & 20.461 $\pm$ 0.028 & 19.937 $\pm$ 0.026 & 20.520 $\pm$ 0.045 & 21.273 $\pm$ 0.034 & 20.586 $\pm$ 0.027 & 20.235 $\pm$ 0.025\\
WISEA J163940.84$-$684739.4 & Y0pec & 20.833 $\pm$ 0.023 & 20.626 $\pm$ 0.023 & 20.746 $\pm$ 0.029 & 21.674 $\pm$ 0.025 & 21.252 $\pm$ 0.024 & 20.824 $\pm$ 0.024\\
WISEA J173835.52$+$273258.8\tablenotemark{a} & Y0 & \dots & 19.546 $\pm$ 0.023 & 20.246 $\pm$ 0.031 & \dots & 20.223 $\pm$ 0.023 & 19.923 $\pm$ 0.023\\
WISEA J205628.88$+$145953.6\tablenotemark{a} & Y0 & \dots & 19.129 $\pm$ 0.022 & 19.643 $\pm$ 0.026 & \dots & 19.811 $\pm$ 0.023 & 19.479 $\pm$ 0.022\\
WISEA J220905.75$+$271143.6 & Y0: & 22.954 $\pm$ 0.071 & 22.859 $\pm$ 0.128 & 22.389 $\pm$ 0.152 & 23.508 $\pm$ 0.110 & 23.355 $\pm$ 0.156 & 22.877 $\pm$ 0.105\\
WISEA J221216.27$-$693121.6 & T9 & 20.282 $\pm$ 0.023 & 19.737 $\pm$ 0.024 & 20.225 $\pm$ 0.036 & 21.043 $\pm$ 0.027 & 20.378 $\pm$ 0.026 & 20.047 $\pm$ 0.024\\
WISEA J222055.34$-$362817.5 & Y0 & 20.899 $\pm$ 0.034 & 20.447 $\pm$ 0.025 & 20.858 $\pm$ 0.035 & 21.783 $\pm$ 0.042 & 21.131 $\pm$ 0.027 & 20.749 $\pm$ 0.024\\
WISEA J235402.79$+$024014.1\tablenotemark{a} & Y1 & \dots & 23.068 $\pm$ 0.199 & 22.882 $\pm$ 0.300 & \dots & 24.124 $\pm$ 0.393 & 23.292 $\pm$ 0.179
\enddata
\tablenotetext{a}{Y-band and F105W photometry were not synthesized for these dwarfs because they were not observed with the G102 grism.}
\tablenotetext{b}{H-band and F140W photometry were not synthesized for WISE 0535$-$7500 because its spectrum is contaminated in this wavelength range.}
\\
\end{deluxetable}

\clearpage

\begin{deluxetable}{cccccccccl}
\tablecaption{Models Used for Spectral Fitting$^a$}
\tabletypesize{\scriptsize}
\tablewidth{0pt}
\tablehead{T$_{eff}$ & log $g$ & $f_{sed}$ & Cloud Type & Reference\tablenotemark{ } \\
(K) & (cm s$^{-1}$) &  & (\%)}
\startdata
400$-$700(50), 700$-$1000(100) & 3.0 & nc & \dots & 1\\
400$-$1000(50) & 3.5 & nc & \dots & 1\\
300$-$1000(50) & 3.75 & nc & \dots & 1\\
300$-$1000(50) & 4.0 & nc & \dots & 1\\
400$-$1000(50) & 4.25 & nc & \dots & 1\\
300$-$1000(50) & 4.5 & nc & \dots & 1\\
400$-$1000(50) & 4.75 & nc & \dots & 1\\
400$-$1000(50) & 5.0 & nc & \dots & 1\\
400$-$1000(100) & 5.5 & nc & \dots & 1\\
200$-$400(25) & 3.0 & 5 & 50\tablenotemark{b} & 2\\
200$-$400(25), 450 & 3.5 & 5 & 50\tablenotemark{b} & 2\\
200$-$400(25), 450 & 4.0 & 3 & 50\tablenotemark{b} & 2\\
200$-$400(25), 450 & 4.0 & 5 & 50\tablenotemark{b} & 2\\
200$-$400(25), 450 & 4.0 & 7 & 50\tablenotemark{b} & 2\\
200$-$400(25), 450 & 4.5 & 5 & 50\tablenotemark{b} & 2\\
200$-$400(25), 400$-$700(50), 700$-$1000(100)  & 5.0 & 5 & 50\tablenotemark{b} & 2\\
400, 450, 550, 600$-$1000(100) & 4.0 & 2 & 100 & 3\\
400$-$600(50), 600$-$1000(100) & 4.0 & 3 & 100 & 3\\
400$-$600(50), 600$-$1000(100) & 4.0 & 4 & 100 & 3\\
400$-$600(50), 600$-$1000(100) & 4.0 & 5 & 100 & 3\\
400$-$600(50), 600$-$1000(100) & 4.5 & 2 & 100 & 3\\
400$-$600(50), 600$-$1000(100) & 4.5 & 3 & 100 & 3\\
400$-$600(50), 600$-$1000(100) & 4.5 & 4 & 100 & 3\\
400$-$600(50), 600$-$1000(100) & 4.5 & 5 & 100 & 3\\
460, 500$-$600(50), 600$-$1000(100) & 5.0 & 2 & 100 & 3\\
400$-$600(50), 460, 600$-$1000(100) & 5.0 & 3 & 100 & 3\\
400, 460, 500$-$600(50), 600$-$1000(100) & 5.0 & 4 & 100 & 3\\
400, 460, 500$-$600(50), 600$-$1000(100) & 5.0 & 5 & 100 & 3\\
400, 500$-$600(50), 600$-$ 1000(100) & 5.5 & 3 & 100 & 3
\enddata
\tablenotetext{ }{References: (1) \citealt{sau12}; (2) \citealt{mor14}; (3) \citealt{mor12}}
\tablenotetext{a}{When a range is given, the value given in parentheses is the spacing for that interval.}
\tablenotetext{b}{With water clouds.}
\end{deluxetable}
\clearpage

\begin{deluxetable}{lccccccccl}
\tablecaption{Model-derived Physical Characteristics\tablenotemark{a}}
\tabletypesize{\scriptsize}
\tablewidth{0pt}
\tablehead{
\colhead{AllWISE Name} & \colhead{Spec.} & T$_{eff}$ & log $g$ & $f_{sed}$ & \% Cloudy & (R/d) & $\chi^2$/d.o.f & Age\tablenotemark{b} \\
& Type & (K) & (cm s$^{-1}$) &  & (\%) & (R$_{jup}$/pc) & & (Gyr)  } 
\startdata
WISEA J032504.52$-$504403.0 & T8 & 550$-$600 & 4.0 & 3$-$4 & 100 & 2.468$\times$10$^{-2}$ & 5213/211 & 0.08$-$0.3 \\
WISEA J033515.07$+$431044.7 & T9 & 500$-$550 & 4.0$-$4.5 & 3$-$5 & 100 & 1.973$\times$10$^{-2}$ & 5276/211 & 0.2$-$1.5 \\
WISEA J035000.31$-$565830.5 & Y1 & 300$-$350 & 5.0 & \dots,5 & 0,50 & 4.704$\times$10$^{-2}$ & 2873/128 & U \\
WISEA J035934.07$-$540154.8 & Y0 & 400 & 4.0$-$5.0 & 2,3,5 & 100 & 3.641$\times$10$^{-2}$ & 2494/128 & 1.5$-$8 \\
WISEA J040443.50$-$642030.0 & T9 & 550$-$600 & 4.0 & 3$-$4 & 100 & 1.861$\times$10$^{-2}$ & 4458/211 & 0.08$-$0.3 \\
WISEA J041022.75$+$150247.9 & Y0 & 400 & 4.0$-$4.5 & 2,3,4 & 100 & 1.007$\times$10$^{-1}$ & 5172/128 & 1.5$-$8 \\
WISEA J053516.87$-$750024.6 & $\geq$Y1 & 450$-$500 & 5.0 & \dots,5 & 0,50 & 8.397$\times$10$^{-3}$ & 2744/179 & 4$-$U \\
WISEA J064723.24$-$623235.4 & Y1 & 350$-$400 & 5.0 & \dots,5 & 0,50 & 2.609$\times$10$^{-2}$ & 2621/211 & 8$-$U \\
WISEA J073444.03$-$715743.8 & Y0 & 450 & 4.0$-$4.5 & 3$-$5 & 100 & 2.410$\times$10$^{-2}$ & 3109/211 & 0.4$-$2 \\
WISEA J082507.37$+$280548.2 & Y0.5 & 400 & 4.5$-$5.0 & \dots,4,5 & 0,100 & 1.808$\times$10$^{-2}$ & 3313/211 & 3$-$U \\
WISEA J094306.00$+$360723.3 & T9.5 & 450$-$500 & 4.0$-$4.5 & 2,3,5 & 100 & 4.167$\times$10$^{-2}$ & 4506/128 & 0.3$-$2 \\
WISEA J120604.25$+$840110.5 & Y0 & 400$-$450 & 4.0$-$4.5 & 3,5 & 100 & 4.040$\times$10$^{-2}$ & 3475/211 & 0.4$-$3 \\
WISEA J140518.32$+$553421.3 & Y0.5 & 350$-$400 & 5.0$-$5.5 & \dots,5 & 0,100 & 3.218$\times$10$^{-2}$ & 4117/211 & U \\
WISE J154151.65$-$225024.9 & Y1 & 400 & 4.0$-$4.5 & \dots & 0 & 2.072$\times$10$^{-2}$ & 4102/211 & 0.6$-$3 \\
WISEA J154214.00$+$223005.2 & T9.5 & 450-500 & 4.0$-$4.5 & 3$-$5 & 100 & 2.127$\times$10$^{-2}$ & 4051/211 & 0.3$-$2 \\
WISEA J163940.84$-$684739.4 & Y0pec & 400 & 5.0 & \dots,5 & 0,100 & 3.186$\times$10$^{-2}$ & 10859/211 & 8$-$U \\
WISEA J173835.52$+$273258.8 & Y0 & 400 & 4.0$-$4.5 & 2,4,5 & 100 & 6.445$\times$10$^{-2}$ & 6598/128 & 0.6$-$3 \\
WISEA J205628.88$+$145953.6 & Y0 & 400$-$450 & 4.0$-$4.5 & 2,4 & 100 & 5.541$\times$10$^{-2}$ & 10770/128 & 0.4$-$3 \\
WISEA J220905.75$+$271143.6 & Y0: & 500$-$550 & 4.0$-$4.5 & 4,5 & 100 & 6.682$\times$10$^{-3}$ & 3106/211 & 0.2$-$1.5 \\
WISEA J221216.27$-$693121.6 & T9 & 500$-$600 & 4.0 & 3,5 & 100 & 1.541$\times$10$^{-2}$ & 5389/211 & 0.08$-$0.4 \\
WISEA J222055.34$-$362817.5 & Y0 & 400$-$450 & 4.0$-$5.0 & 2,3,5 & 100 & 3.204$\times$10$^{-2}$ & 4574/211 & 1$-$6 \\
WISEA J235402.79$+$024014.1 & Y1 & 300$-$400 & 4.0$-$5.0 & 2,5 & 50,100  & 5.961$\times$10$^{-2}$ & 10663/128 & 1.5$-$U 
\enddata
\tablenotetext{a}{Model-derived parameters are generally unreliable for reasons explained in Section 5.4}
\tablenotetext{b}{Ages are derived for the T$_{eff}$ and log $g$ ranges in columns 3 and 4 using the \cite{sau08} cloudless evolutionary models.  Some combinations of T$_{eff}$ and log $g$ derived from the atmospheric fits are unphysical in that brown dwarfs cannot evolve to have such values, so the corresponding ages are denoted as `U' for unphysical.}

\end{deluxetable}

\end{center}


\begin{thebibliography}{}
\bibitem[Beichman et al.(2014)]{beich14} Beichman, C., Gelino, C.~R., Kirkpatrick, J.~D., et al.\ 2014, \apj, 783, 68
\bibitem[Burgasser et al.(2006)]{burg06} Burgasser, A.~J., Geballe, T.~R., Leggett, S.~K., Kirkpatrick, J.~D., \& Golimowski, D.~A.\ 2006, \apj, 637, 1067
\bibitem[Burrows et al.(2000)]{bur00} Burrows, A., Marley, M.~S., \& Sharp, C.~M.\ 2000, \apj, 531, 438 
\bibitem[Burrows et al.(2003)]{bur03} Burrows, A., Sudarsky, D., \& Lunine, J.~I.\ 2003, \apj, 596, 587
\bibitem[Cushing et al.(2006)]{cush06} Cushing, M.~C., Roellig, T.~L., Marley, M.~S., et al.\ 2006, \apj, 648, 614 
\bibitem[Cushing et al.(2008)]{cush08} Cushing, M.~C., Marley, M.~S., Saumon, D., et al.\ 2008, \apj, 678, 1372 
\bibitem[Cushing et al.(2011)]{cush11} Cushing, M.~C., Kirkpatrick, J.~D., Gelino, C.~R., et al.\ 2011, \apj, 743, 50
\bibitem[Cushing et al.(2014a)]{cush14a} Cushing, M.~C., Kirkpatrick, J.~D., Gelino, C.~R., et al.\ 2014a, \aj, 147, 113
\bibitem[Cutri et al.(2003)]{cut03} Cutri, R.~M., Skrutskie, M.~F., van Dyk, S., et al. 2003, VizieR Online Data Catalog, 2246, 0
\bibitem[Cutri et al.(2012)]{cut12} Cutri, R.~M., et al. 2012, VizieR Online Data Catalog, 2311, 0
\bibitem[Dupuy \& Liu(2012)]{dup12} Dupuy, T.~J., \& Liu, M.~C.\ 2012, \apjs, 201, 19
\bibitem[Dupuy \& Kraus(2013)]{dup13} Dupuy, T.~J., \& Kraus, A.~L.\ 2013, Science, 341, 1492
\bibitem[Faherty et al.(2014)]{fah14} Faherty, J.~K., Tinney, C.~G., Skemer, A., \& Monson, A.~J.\ 2014, \apjl, 793, LL16 
\bibitem[Fazio et al.(2004)]{faz04} Fazio, G.~G., Hora, J.~L., Allen, L.~E., et al.\ 2004, \apjs, 154, 10 
\bibitem[Golimowski et al.(2004)]{gol04} Golimowski, D.~A., Leggett, S.~K., Marley, M.~S., et al.\ 2004, \aj, 127, 3516 
\bibitem[Gonzaga et al.(2012)]{gonz12} Gonzaga, S., \& et al.\ 2012, The DrizzlePac Handbook, HST Data Handbook, (Baltimore, STScI)
\bibitem[Hubeny \& Burrows(2007)]{hub07} Hubeny, I., \& Burrows, A.\ 2007, \apj, 669, 1248 
\bibitem[Kimble et al.(2008)]{kim08} Kimble, R.~A., MacKenty, J.~W., O'Connell, R.~W., \& Townsend, J.~A.\ 2008, \procspie, 7010, 43
\bibitem[Kirkpatrick(2008)]{kirk08} Kirkpatrick, J.~D.\ 2008, 14th Cambridge Workshop on Cool Stars, Stellar Systems, and the Sun, 384, 85 
\bibitem[Kirkpatrick et al.(2011)]{kirk11} Kirkpatrick, J.~D., Cushing, M.~C., Gelino, C.~R., et al.\ 2011, \apjs, 197, 19
\bibitem[Kirkpatrick et al.(2012)]{kirk12} Kirkpatrick, J.~D., Gelino, C.~R., Cushing, M.~C., et al.\ 2012, \apj, 753, 156
\bibitem[Kirkpatrick et al.(2013)]{kirk13} Kirkpatrick, J.~D., Cushing, M.~C., Gelino, C.~R., et al.\ 2013, \apj, 776, 128
\bibitem[Kirkpatrick et al.(2014)]{kirk14} Kirkpatrick, J.~D., Schneider, A., Fajardo-Acosta, S., et al.\ 2014, \apj, 783, 122 
\bibitem[Kuntschner et al.(2011)]{kun11} Kuntschner, H., K{\"u}mmel, M., Walsh, J.~R., \& Bushouse, H.\ 2011, Space Telescope WFC Instrument Science Report, 5 
\bibitem[Leggett et al.(2007)]{leg07} Leggett, S.~K., Marley, M.~S., Freedman, R., et al.\ 2007, \apj, 667, 537
\bibitem[Leggett et al.(2010)]{leg10} Leggett, S.~K., Burningham, B., Saumon, D., et al.\ 2010, \apj, 710, 1627
\bibitem[Leggett et al.(2013)]{leg13} Leggett, S.~K., Morley, C.~V., Marley, M.~S., et al.\ 2013, \apj, 763, 130
\bibitem[Leggett et al.(2014a)]{leg14} Leggett, S.~K., Liu, M.~C., Dupuy, T.~J., et al.\ 2014a, \apj, 780, 62
\bibitem[Leggett et al.(2014b)]{leg14b} Leggett, S.~K.,  Morley, C.~V., Marley, M.~S., \& Saumon, D.\ 2014b, arXiv:1404.2020
\bibitem[Liu et al.(2011)]{liu11} Liu, M.~C., Delorme, P., Dupuy, T.~J., et al.\ 2011, \apj, 740, 108 
\bibitem[Liu et al.(2012)]{liu12} Liu, M.~C., Dupuy, T.~J., Bowler, B.~P., Leggett, S.~K., \& Best, W.~M.~J.\ 2012, \apj, 758, 57
\bibitem[Lodieu et al.(2013)]{lod13} Lodieu, N., B{\'e}jar, V.~J.~S., \& Rebolo, R.\ 2013, \aap, 550, L2 
\bibitem[Luhman et al.(2011)]{luh11} Luhman, K.~L., Burgasser, A.~J., \& Bochanski, J.~J.\ 2011, \apjl, 730, L9 
\bibitem[Luhman(2014)]{luh14} Luhman, K.~L.\ 2014, \apjl, 786, L18 
\bibitem[Luhman \& Esplin(2014)]{luh14b} Luhman, K.~L., \& Esplin, T.~L.\ 2014, \apj, 796, 6
\bibitem[Mace et al.(2013)]{mace13} Mace, G.~N., Kirkpatrick, J.~D., Cushing, M.~C., et al.\ 2013, \apjs, 205, 6
\bibitem[Morley et al.(2012)]{mor12} Morley, C.~V., Fortney, J.~J., Marley, M.~S., et al.\ 2012, \apj, 756, 172 
\bibitem[Morley et al.(2014)]{mor14} Morley, C.~V., Marley, M.~S., Fortney, J.~J., et al.\ 2014, \apj, 787, 78 
\bibitem[Pinfield et al.(2014)]{pin14} Pinfield, D.~J., Gromadzki, M., Leggett, S.~K., et al.\ 2014, \mnras, 444, 1931 
\bibitem[Saumon et al.(2003)]{sau03} Saumon, D., Marley, M.~S., Lodders, K., \& Freedman, R.~S.\ 2003, Brown Dwarfs, 211, 345 
\bibitem[Saumon \& Marley(2008)]{sau08} Saumon, D., \& Marley, M.~S.\ 2008, \apj, 689, 1327 
\bibitem[Saumon et al.(2012)]{sau12} Saumon, D., Marley, M.~S., Abel, M., Frommhold, L., \& Freedman, R.~S.\ 2012, \apj, 750, 74
\bibitem[Schneider et al.(2014)]{sch14} Schneider, A.~C., Cushing, M.~C., Kirkpatrick, J.~D., et al.\ 2014, \aj, 147, 34 
\bibitem[Stephens et al.(2009)]{ste09} Stephens, D.~C., Leggett, S.~K., Cushing, M.~C., et al.\ 2009, \apj, 702, 154 
\bibitem[Thompson et al.(2013)]{thom13} Thompson, M.~A., Kirkpatrick, J.~D., Mace, G.~N., et al.\ 2013, \pasp, 125, 809
\bibitem[Tinney et al.(2012)]{tin12} Tinney, C.~G., Faherty, J.~K., Kirkpatrick, J.~D., et al.\ 2012, \apj, 759, 60
\bibitem[Tokunaga \& Vacca(2005)]{tok05} Tokunaga, A.~T., \& Vacca, W.~D.\ 2005, \pasp, 117, 421 
\bibitem[van Dokkum(2001)]{vand01} van Dokkum, P.~G.\ 2001, \pasp, 113, 1420 
\bibitem[Vrba et al.(2004)]{vrba04} Vrba, F.~J., Henden, A.~A., Luginbuhl, C.~B., et al.\ 2004, \aj, 127, 2948 
\bibitem[Witte et al.(2011)]{witte11} Witte, S., Helling, C., Barman, T., Heidrich, N., \& Hauschildt, P.~H.\ 2011, \aap, 529, AA44 
\bibitem[Wright et al.(2010)]{wri10} Wright, E.~L., Eisenhardt, P.~R.~M., Mainzer, A.~K., et al.\ 2010, \aj, 140, 1868
\end{thebibliography}
\end{document}